\documentclass[11pt]{article}

\usepackage{geometry,graphicx,amsmath,amssymb,amsfonts,mathdots,mathrsfs,caption,subcaption, setspace, wrapfig}
\usepackage[usenames,dvipsnames,svgnames,table]{xcolor}
\usepackage{graphicx}
\usepackage{jheppub}
\usepackage{epsfig}
\usepackage{longtable}
\usepackage{braket}
\usepackage{comment}
\usepackage{slashed}

\usepackage{ulem}

\newcommand{\be}{\begin{equation}}
\newcommand{\ee}{\end{equation}}
\newcommand{\bea}{\begin{eqnarray}}
\newcommand{\eea}{\end{eqnarray}}
\newcommand{\bwt}{\begin{widetext}}
\newcommand{\ewt}{\end{widetext}}

\begin{document}
\title{Search for Heavy Right-Handed Neutrinos at the LHC and Beyond in the Same-Sign Same-Flavor Leptons Final State}

\author{John N. Ng$^{1,}$\footnote{misery@triumf.ca}, Alejandro de la Puente$^{1,}$\footnote{adelapue@triumf.ca}, and Bob Wei-Ping Pan$^{2,}$\footnote{two.joker@gmail.com}
\vspace{5mm}
\\
\normalsize\emph{(1) Theory Group, TRIUMF,} \\
\normalsize\emph{4004 Wesbrook Mall}\\
\normalsize\emph{Vancouver BC V6T 2A3, Canada.} \\
\normalsize\emph{(2) Department of Physics } \\
\normalsize\emph{National Tsing Hua Univeristy} \\
\normalsize\emph{Hsin Chu 300, Taiwan, Republic of China}\\}
\date{\today}

\abstract{

In this study we explore the LHC's Run II potential to the discovery of heavy Majorana neutrinos, with luminosities between $30$ and $3000$ fb$^{-1}$ in the $l^{\pm}l^{\pm}j~j$ final state. Given that there exist many models for neutrino mass generation, even within the Type I seesaw framework, we use a simplified model approach and study two simple extensions to the Standard Model, one with a single heavy Majorana neutrino, singlet under the Standard Model gauge group, and a limiting case of the left-right symmetric model. We then extend the analysis to a future hadron collider running at $100$ TeV center of mass energies. This extrapolation in energy allows us to study the relative importance of the resonant production versus gauge boson fusion processes in the study of Majorana neutrinos at hadron colliders. We analyze and propose different search strategies designed to maximize the discovery potential in either the resonant production or the gauge boson fusion modes.

}

\maketitle

\section{Introduction}

There is convincing evidence for the existence of three active neutrino species~\cite{Ade:2013zuv,ALEPH:2005ab}, at least two are massive. Their mass pattern has been narrowed down to a normal or an inverted hierarchy or degenerate~\cite{Homestake,Gallex,Sage,Boxerino,SuperK,SNO}; however, the absolute mass scale remains unknown. In addition, we are now capable of measuring CP violation in the lepton sector, given that the last mixing angle, $\theta_{13}$, has been measured by several reactor experiments~\cite{Chooz,Daya,Reno} and the T2K accelerator experiment~\cite{T2K}. The experimental status of neutrino physics suggests that the evidence for neutrino masses represents a clear motivation for new physics beyond the Standard Model (SM). In principle this is due to the fact that within the SM, neutrinos are massless and one may incorporate new degrees of freedom or effective interactions to generate Dirac or Majorana masses. The latter is an interesting possibility since a Majorana mass term violates lepton number by two units. While a neutrinoless double beta decay ($0\nu\beta\beta$) signal would be a breakthrough giving us knowledge on the nature of neutrinos and with additional assumptions the scale of the active neutrino mass matrix, the observation of lepton flavor violating processes at the LHC can shake the foundations of the SM with additional mechanisms and with luck we may directly find the new particles associated with these interactions.

In the minimal Type I seesaw mechanism~\cite{Seesaw}, the SM is extended with a single Majorana fermion, singlet under the SM gauge group. Within this framework, the Majorana fermion carries lepton number and couples to left-handed leptons through the Higgs field. In addition, a Majorana mass term for the Majorana fermion can be implemented consistent with the gauge symmetries of the SM. The Majorana nature of this fermion and its mixing with the SM active neutrinos may lead to interesting testable interactions at the LHC that violate lepton number such as the production of same-sign leptons in association with jets. This final state can be achieved through resonant production of a Majorana neutrino and through gauge boson fusion. There has been a number of theoretical studies aimed at examining the sensitivity that the LHC has to heavy Majorana neutrino masses and to set limits on the couplings to leptons that arise from mixing with the active SM neutrinos~\cite{Datta:1993nm,Almeida:2000pz,Panella:2001wq,Han:2006ip,del Aguila:2007em,Atre:2009rg,Dev:2013wba,Deppisch:2015qwa}. In particular, a 14 TeV LHC with $100$ fb$^{-1}$ may have sensitivity to resonant production of Majorana neutrinos with masses up to $\sim400$ GeV using the $\mu^{\pm}\mu^{\pm}$ and $\mu^{\pm}e^{\pm}$ channels. However, the $W\gamma$ fusion can reach masses up to $530$ GeV in the $\mu^{\pm}\mu^{\pm}$ channel~\cite{Alva:2014gxa}. Furthermore, in the regime where the Majorana Neutrino has mass below the mass of the SM $W$ boson, the authors in~\cite{Izaguirre:2015pga} have shown that the LHC can be sensitive to mixing angles in the range $10^{-4}-10^{-3}$ and Majorana neutrino masses below $15$ GeV. This is done by looking for a prompt lepton in association with a displaced lepton jet. For heavy Majorana masses, no significant reach can be achieved with the gauge bosons fusion process.  These channels are  collider analogue to ($0\nu\beta\beta$) which are not available there. Moreover, the $ee$ channel may be sensitive to electron-like Majorana neutrinos with masses between $150-1400$ GeV and higher LHC energies~\cite{Dicus:1991fk}.

Alternatively, one can extend the SM to be left-right symmetric; partnering right-handed charged leptons with new right-handed neutrinos into an SU(2)$_{\text{R}}$ doublet. Within this class of models, the gauge couplings are left-right symmetric and a new charged gauge boson, $W_{R}$, connects right-handed charge currents~\cite{Pati:1974yy,Mohapatra:1974hk,Mohapatra:1974gc,Senjanovic:1975rk,Senjanovic:1978ev}. The phenomenology of left-right symmetric models is very rich with the appearance of new scalar degrees of freedom and lepton violating process such as the production of same-sign leptons. The latter can be induced from the production of a right-handed charged gauge boson. A number of theoretical studies have examined the sensitivity that the LHC has to elements of the right-handed lepton mixing matrix, the mass of $W_{R}$ and Majorana neutrino masses~\cite{Keung:1983uu,Tello:2010am,Chen:2011hc,Aguilar-Saavedra:2014ola,Heikinheimo:2014tba,Deppisch:2014qpa,Vasquez:2014mxa,Das:2012ii,Han:2012vk,Chakrabortty:2012pp,Chen:2013foz,Gluza:2015goa}.

With the LHC running, both CMS and ATLAS have searched for same-sign dilepton final states using light- or $b$-jets~\cite{Chatrchyan:2013fea,Chatrchyan:2012paa,TheATLAScollaboration:2013jha}. In addition, a recent analysis by CMS~\cite{Khachatryan:2015gha} is used to search for heavy Majorana neutrinos in the same-sign dimuon channel with an integrated luminosity of $19.7$ fb$^{-1}$ and 8 TeV center of mass energies with a sensitivity to a Majorana mass of $500$ GeV. Dedicated searches to probe same-sign leptons within the framework of left-right symmetric models have also been carried out. In particular, both CMS and ATLAS have set limits on heavy Majorana neutrino production assuming $M_{W_{R}}>M_{N_{R}}$ and identical quark and neutrino mixing matrices for the left- and right-handed interactions~\cite{Khachatryan:2014dka,ATLAS:2012ak}. The exclusion regions extending to $M_{W_{R}}\sim3$ TeV in the ($M_{W_{R}},M_{N_{R}})$ plane.

Our work compliments and extends the various analyses discussed above. In the present analysis we explore the LHC's Run II potential to discover heavy Majorana neutrinos with luminosities between $30$ and $3000$ fb$^{-1}$ the $l^{\pm}l^{\pm}j~j$ final state. Given that there exist many models for neutrino mass generation, even within the Type I seesaw framework, we use a simplified model approach and study two simple extensions to the Standard Model, one with a single heavy Majorana neutrino, singlet under the Standard Model gauge group, and a limiting case of the left-right symmetric model. We would like to emphasize that simplified models are not complete models. They are constructed
to highlight specific points which in our case means LHC signals, and they can be mapped into more realistic models. They can also be viewed as truncations of more complete models. For the Type I Seesaw, a more natural framework would include at least two heavy right-handed neutrinos to explain the mass differences between the active neutrinos. However, in the mass eigenstate basis of these new heavy Majorana neutrinos the mixing angles and the masses can be treated as independent variables. Furthermore, more details regarding the model are necessary to determine the correlation between the new parameters and the mass differences of the active neutrinos. We will be conservative and assume that only the lightest of the new heavy sates are within reach of the LHC or future circular collider.
Incorporating a second and third heavy Majorana neutrino simply leads to a duplication of our results; thus we ignore the heavier degrees of freedom and also the mass relation between active neutrinos. Thus, we can focus on the lightest heavy Majorana neutrino without loss of generality. Furthermore, the usual Type I seesaw model has very small mixing of light and heavy neutrinos. This can be circumvented by arranging structures in the heavy neutrino mass matrix. The details are highly model dependent; albeit important for fitting low energy neutrino data, see e.g.~\cite{ZhZh,Ibarra:2010xw}. In the simplified model this is captured by allowing the couplings of the heavy neutrino to the charged leptons to be free parameters.
 We then extend the analysis to a future hadron collider running at $100$ TeV center of mass energies. This extrapolation in energy allows us to study the relative importance of the resonant production versus gauge boson fusion processes in the study of Majorana neutrinos at hadron colliders. We analyze and propose different search strategies designed to maximize the discovery potential in either the resonant production or the gauge boson fusion modes. Our work is strictly phenomenological.
We check consistency with the strongest model independent constraints that arise from low energy interactions such as rare decays of the muon, unitarity of the PMNS matrix, and the null evidence for ($0\nu\beta\beta$). A well known caveat to be added here is that other than unitarity tests, low energy constraints are obtained with the assumption
that other new physics do not contribute significantly.

The summary of our study is as follows: In Section~\ref{sec:model} we introduce the models and interpret current low energy and collider constraints. In Section~\ref{sec:Signal} we calculate the leading SM backgrounds and propose various search strategies designed to maximize the discovery potential in either the resonant production or the gauge boson fusion modes of a heavy Majorana neutrino. In Section~\ref{sec:discussion} we provide concluding remarks.

\section{Two Simplified Models}\label{sec:model}
To study same-sign leptons production in association with jets at the LHC run II and future higher energy colliders, we examine two simplified models that give rise to this final state topology. The first one consists only of SM gauge interactions and the only added new degree of freedom is the singlet heavy Majorana $N_R$ fermion which
mixes with the 3 active SM neutrinos. The mass of  $N_R$ and its mixing with the active neutrinos are taken to be free parameters.
  The second model implements an additional SU$(2)$ gauge symmetry under which the right-handed leptons and quarks are charged unlike the left-handed chirality. $N_R$ now partners with the right-handed charged leptons. This can be part of a left-right symmetric model that can give rise to realistic masses for the active neutrinos. The details are not essential for our purpose. In what follows, we review different aspect of these two theoretical approaches and indicate how they can be related to more complete models.

\subsection{Single Majorana Fermion Extension to the Standard Model}\label{subsec:onlySM}

The following framework consists of an extension to the SM with a single Majorana fermion, singlet under the SM gauge interactions. Within this analysis, the Majorana fermion, $N_{R}$, mixes with the three active neutrinos. We are aware that to get realistic active neutrino masses, two or more Majorana neutrinos have to be added. In the spirit of the simplified model we are considering we can assume that these additional states are either much heavier
or have negligible mixings.  Following the analysis detailed in~\cite{Han:2006ip} where in terms of mass eigenstates, the gauge interaction Lagrangian is given by:
\begin{eqnarray}
{\cal L}&=&-\frac{g}{\sqrt{2}}W^{+}_{\mu}\left(\sum_{l=e}^{\tau}\sum_{m=1}^{3}U^{*}_{lm}\bar{\nu}_{m}\gamma^{\mu}P_{L}l\right)-\frac{g}{\sqrt{2}}W^{+}_{\mu}\left(\sum_{l=e}^{\tau}V^{*}_{l4}N^{c}_{R}\gamma^{\mu}P_{L}l\right) \nonumber \\
&-&\frac{g}{2\cos\theta_{W}}Z_{\mu}\left(\sum_{l=e}^{\tau}V^{*}_{l4}N^{c}_{R}\gamma^{\mu}P_{L}\nu_{l}\right)+~\text{H.c},\label{eq:ccSM}
\end{eqnarray}
where $U$ and $V$ provide a relationship between the flavor basis, $\nu_{lL}'$ and the mass basis parametrized by $\nu_{mL}$ such that
\begin{equation}
\nu'_{lL}=\sum_{m=1}^{3}U_{lm}\nu_{m}+V_{lN}N^{c}_{L},
\end{equation}
with $UU^{\dagger}+VV^{\dagger}=1$. We refer the reader to~\cite{Atre:2009rg} for more details regarding the mixing formalism. In our simplified approach, we parametrize the charged current interactions by introducing the parameters $\epsilon_{l}$, where
\begin{equation}
\epsilon_{l}=V_{lN}.
\end{equation}
For Type I seesaw models the mass of the Majorana neutrino is commonly assumed to be $M_{N_{R}}\ge10^{14}$ GeV to obtain active neutrino masses below $2$ eV~\cite{AbsScale}. However, there exist well motivated models that can lower the seesaw scale and/or have a sizable coupling to the SM particles. Notable examples are structures or cancelations in the heavy neutrino sector that can lead to $M_{N_{R}}\sim {\cal O}\left(\text{TeV}\right)$ and larger mixings~\cite{Ibarra:2010xw}. Approximately conserved lepton number has also been invoked~\cite{GonzalezGarcia:1988rw,Kersten:2007vk}. The inverse seesaw mechanism can also yield TeV scale $N_{R}$ masses with mixings at a few percent~\cite{Chang:2013yva}. In our analysis we will take a model independent approach and vary the $\epsilon$ couplings freely to investigate the range that the LHC can probe.
\begin{figure}[ht]
\begin{center}
\includegraphics[width=3.0in]{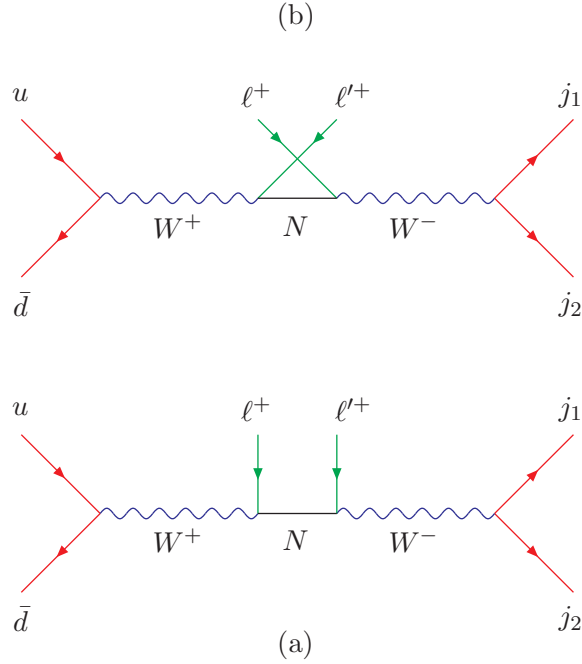}
\caption{\small Feynman diagrams for the $S$-channel contribution to same-sign lepton production. }\label{fig:schannel}
\end{center}
\end{figure}

\begin{figure}[ht]\centering
\begin{subfigure}[b]{0.3\textwidth}
\includegraphics[width=\textwidth]{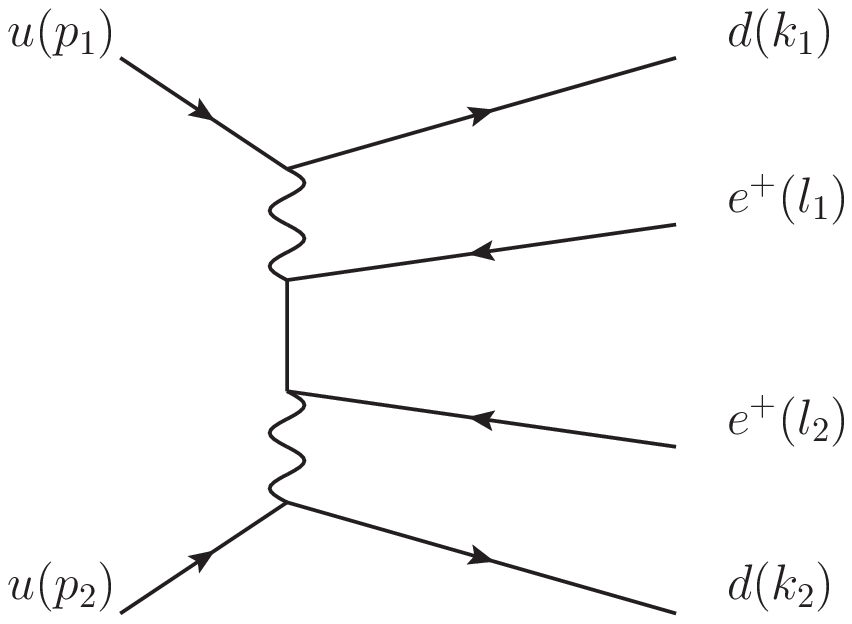}\subcaption{$M_{B}$}
\end{subfigure}
\begin{subfigure}[b]{0.3\textwidth}
\includegraphics[width=\textwidth]{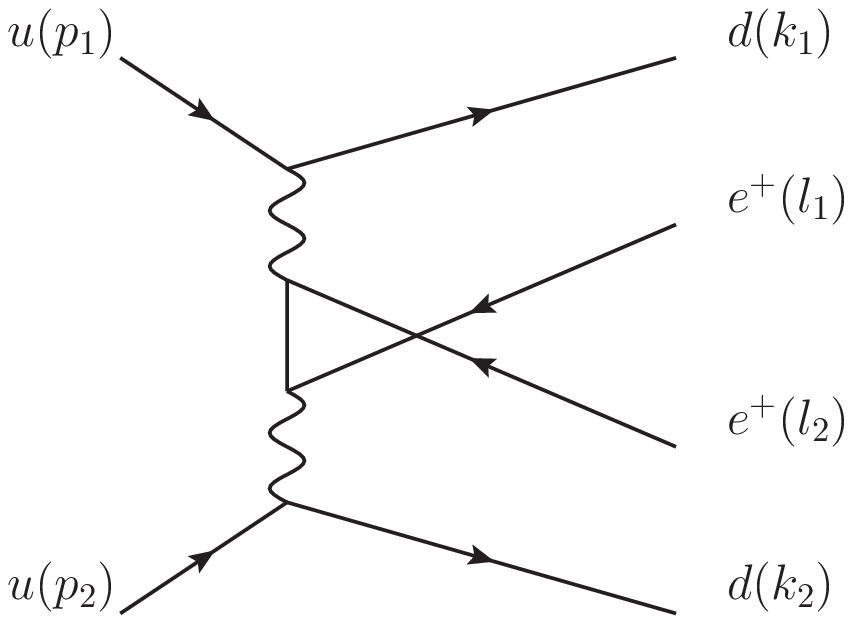}\subcaption{$M_{B}$}
\end{subfigure}
\begin{subfigure}[b]{0.3\textwidth}
\includegraphics[width=\textwidth]{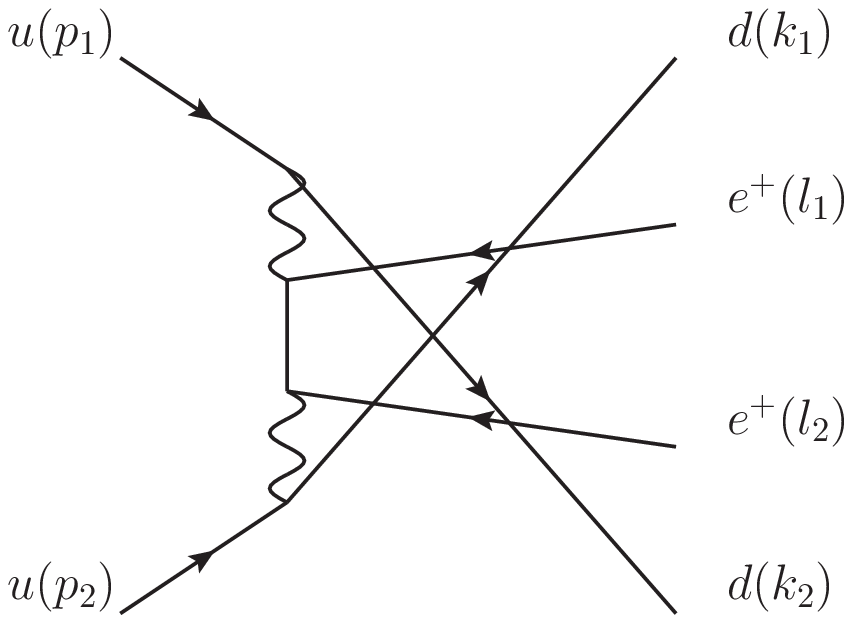}\subcaption{$M_{C}$}
\end{subfigure}
\begin{subfigure}[b]{0.3\textwidth}
\includegraphics[width=\textwidth]{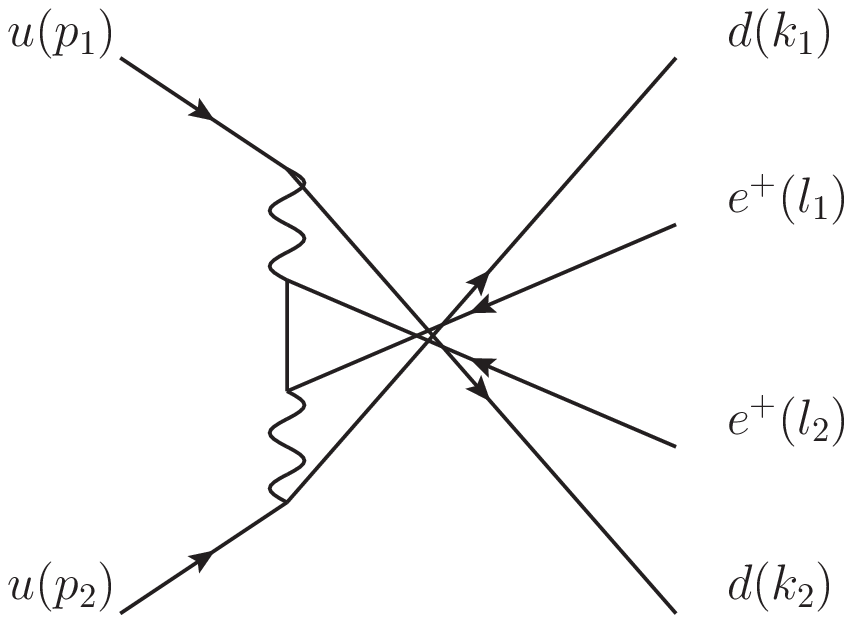}\subcaption{$M_{D}$}
\end{subfigure}
\caption{\small Feynman diagrams for the $T$-channel contribution to same-sign lepton production. Diagrams (c) and (d) correspond to contributions where the final state patrons have the same color.  }\label{fig:WWfusion}
\end{figure}

The signal that will be probed in this study is given by
\begin{equation}
p~p\longrightarrow l^{\pm}~l'^{\pm}~+~j~j\label{eq:SSLeq},
\end{equation}
where $l,l'$ denote either an electron or muon. One important aspect of this reaction is that it contains no missing transverse energy at the parton level. Most of the missing energy will appear after a full detector simulation has been carried out, mostly due to the misidentification of jets. At the parton level there are two channels to consider:
\begin{enumerate}
\item
s-channel $u\bar{d}$ annihilation depicted in Figure~\ref{fig:schannel}:
\begin{equation}
u~+\bar{d}~\longrightarrow W^{*}\longrightarrow l^{+}~+~N_{R}\longrightarrow l^{+}l'^{+}~+~W,
\end{equation}
with the $W$ decaying into light jets. The Majorana neutrino, $N_{R}$ can be on- or off-shell depending on the mass, $M_{N_{R}}$. In this channel, the $W$ from the decay of $N_{R}$ can allow us to control the large SM backgrounds since one may reconstruct the $W$ mass using the two light jets for not too large $M_{N_{R}}$.
\item
t-channel or $WW$ fusion process depicted in Figure~\ref{fig:WWfusion}:
\begin{equation}
u~+u~\longrightarrow d~+~l^{+}~l'^{+}~+~d.
\end{equation}
In this process, the Majorana neutrino is always off-shell. If the initial quarks have the same color, a contribution to the amplitude will arise by interchanging the two forward outgoing jets.
\end{enumerate}


\subsection{Gauge bosons beyond the Standard Model}\label{subsec:LRS}
The signal topology described above is not unique to models that generate mixing between a heavy "right-handed" Majorana neutrino and the three active neutrino species. In fact, models that go beyond the SM and that include additional gauge degrees of freedom can give rise to same-sign lepton final states. Left-right symmetric models are primarily motivated to address the origin of parity violation in low energy physics. It assumes that the Lagrangian is left-right symmetric, with the observed asymmetry, such as in $\beta$-decay, arising from a vacuum expectation not invariant under parity transformations. The SU(2)$_{\text{L}}$$\times$SU(2)$_{\text{R}}$$\times$U(1)$_{\text{B-L}}$$\times$P$_{\text{L}\leftrightarrow \text{R}}$, at low energies, breaks down to the SM gauge group with new particles mediating new interactions arising at higher energies~\cite{Pati:1974yy,Mohapatra:1974hk,Mohapatra:1974gc,Senjanovic:1975rk,Senjanovic:1978ev}, in particular, right-handed charged currents, $W^{\pm}_{R}$. Below we summarize essential ingredients pertaining to left-right symmetric models in light of a same-sign lepton final state at the LHC and refer the reader to~\cite{LeftRightReview} for a full review of the theory.

In our analysis we assume that new charged currents are associated with a gauged SU(2)$_{\text{R}}$ symmetry and introduce three heavy Majorana fermions, $N^{e,\mu,\tau}_{R}$, that together with the right-handed charged leptons transform under the fundamental representation of the new gauge symmetry. The matter content can be written as
\begin{eqnarray}
q_{L}=\begin{pmatrix}
u \\
d \end{pmatrix}_{L}:(2,1,1/3),~~~~~~~~q_{R}=\begin{pmatrix}
u \\
d \end{pmatrix}_{R}:(1,2,1/3), \nonumber \\
\nonumber \\
\nonumber \\
L_{L}=\begin{pmatrix}
\nu \\
l \end{pmatrix}_{L}:(2,1,-1),~~~~~~~~L_{R}=\begin{pmatrix}
N^{l} \\
l \end{pmatrix}_{R}:(1,2,-1).
\end{eqnarray}
Here by symmetry we have three right-handed neutrinos $N^{l}$. Within this framework, the minimal setup consists of one scalar bidoublet and two complex scalar triplets that are given by
\begin{equation}
\Phi=\begin{pmatrix}
\phi^{0}_{1} & \phi^{+}_{2} \\
\phi^{-}_{2} & \phi^{0}_{2}
\end{pmatrix},~~~~~~~~~~~~~~~\Delta_{L,R}=\begin{pmatrix}
\delta^{+}_{L,R}/\sqrt{2} & \delta^{++}_{L,R} \\
\delta^{0}_{L,R} & -\delta^{+}_{L,R}/\sqrt{2}
\end{pmatrix}.
\end{equation}
 This is required to break the gauge symmetries down to U(1)$_{em}$. The scalar phenomenology is very rich and recent studies have been devoted to the possibility that a doubly charged scalar can be detected at the LHC and future more energetic hadron colliders~\cite{Bambhaniya:2015wna,Bambhaniya:2014cia,Mohapatra:2013cia}. In addition, production and decays of charged scalars could help determine the leptonic right-handed mixing matrix at the LHC~\cite{Vasquez:2014mxa}.
 In principle, the heavy gauge bosons will mix with the SM gauge sector, but this mixing is naturally small due the V-A nature of the charged current observed at low energies. In our analysis we explore the limit where the mixing is negligible.

 In the minimal setup, lepton masses are due to the following Yukawa interactions
 \begin{eqnarray}
 {\cal L}_{Y}&=&\bar{L}_{L}\left(Y_{\Phi}\Phi+\tilde{Y}_{\Phi}\tilde{\Phi}\right)L_{R}+\frac{1}{2}\left(L^{T}_{L}Ci\sigma_{2}Y_{\Delta_{L}}\Delta_{L}L_{L}\right) \nonumber \\
 &+&\frac{1}{2}\left(L^{T}_{R}Ci\sigma_{2}Y_{\Delta_{R}}\Delta_{R}L_{R} \right).
 \end{eqnarray}
In the absence of spontaneous CP violation we obtain a Dirac mass term mixing left- and right-handed leptons given by
\begin{equation}
M_{D}=v_{1}Y_{\Phi}+\tilde{Y}_{\Phi}v_{2},
\end{equation}
where $v_{1}$ and $v_{2}$ are the vacuum expectation values ($vev$) of the Higgs bidoublet, $\phi^{0}_{1}$ and $\phi^{0}_{2}$. In addition both left- and right-handed neutrinos acquire a Majorana mass given by
\begin{eqnarray}
M_{\nu}&=&Y_{\Delta_{L}}v_{L}-M^{T}_{D}\frac{1}{M_{N_{R}}}M_{D}, \nonumber \\
M_{N_{R}}&=&Y_{\Delta_{R}}v_{R},
\end{eqnarray}
with $v_{L}=\left<\delta_{L}\right>$ and $v_{R}=\left<\delta_{R}\right>$. In contrast to the Type I seesaw models $v_{L}$ need not be zero at the tree level, although it is expected
to be small compared to the Fermi scale. The details are model dependent. The diagonalization of the mass matrices proceeds as in the Type I seesaw, leading to charged current interactions that in the mass eigenstate basis are given by
\begin{equation}
{\cal L}_{cc}=\frac{g}{\sqrt{2}}\left(\bar{\nu}_{L}V_{L}\slashed{W}_{L}l_{L}+\bar{N}^{l}_{R}V_{R}\slashed{W}_{R}l_{R}\right)+~\text{H.c},\label{eq:WRcc}
\end{equation}
where $V_{L,R}$ are the $3\times 3$ left- and right-handed mixing matrices respectively. In the above Lagrangian, $V_{L}$ and $V_{R}$ have been obtained in the limit where $M_{D}/M_{N_{R}}\ll1$. In this limit $V_{L}\equiv U_{\text{PMNS}}$ and parametrizes the mixing of light neutrinos through charged current interactions leading to neutrino oscillations, while $V_{R}$ parametrizes new phenomena that can be probed at the LHC and through various lepton flavor violating processes~\cite{Das:2012ii,Vasquez:2014mxa}.

The goal of this study is to probe $V_{R}$, in particular the signal topologies in Figures~\ref{fig:schannel} and~\ref{fig:WWfusion} with $W\to W_{R}$ at the LHC with $13$ TeV center of mass energies and at a future $100$ TeV Collider. To accomplish this, we focus only on the structure given in Equation~(\ref{eq:WRcc}) to parametrize the charged currents in the lepton sector and assume that $W_{R}$ couples at leading order to quarks with a hierarchical structure similar to that of the SM CKM matrix. This is done to suppress the bounds on $M_{W_{R}}$ that arise from measurements of CP violating effects of the $K_{L}-K_{S}$ mass difference and $B_{d},B_{s}$ meson parameters~\cite{Beall:1981ze,Barenboim:1996nd,Bertolini:2014sua}. However we direct the reader to~\cite{Zhang:2007fn,Senjanovic:2014pva,Senjanovic:2015yea} where a general study on right-handed quark mixings can be found.


\subsection{Constraints}\label{subsec:constraints}
In the singlet Majorana extension to the SM introduced Section~\ref{subsec:onlySM}, the Majorana neutrino mixes with active neutrinos with a strength proportional to $\epsilon_{l}$ for $l=e,\mu,\tau$. As a result, this framework is sensitive to model independent constraints that arise from lepton unitarity~\cite{Langacker:1988up,Bergmann:1998rg}. These are
\begin{equation}
|\epsilon_{e}|^{2}<0.012,~~|\epsilon_{\mu}|^{2}<0.0096,~~|\epsilon_{\tau}|^{2}<0.016.\label{eq:lepU}
\end{equation}
\begin{figure}[ht]
\begin{center}
\includegraphics[width=2.5in]{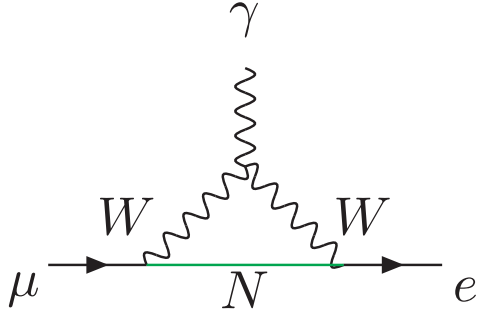}
\caption{\small Contribution to the $\mu\to e\gamma$ decay rate in the presence of a heavy Majorana neutrino. In left-right symmetric models, the the SM $W$ gauge boson can be replaced by $W_{R}$. }\label{fig:mutoeg}
\end{center}
\end{figure}
In addition, indirect constraints that are model dependent arise from rare decays of the muon and muon properties, such as $\mu\to e\gamma$ and the muon anomalous magnetic moment, $a_{\mu}$. The validity of these constraints assumes that $N_{R}$ is the only source of new physics entering the calculation of the matrix elements, such that there is no cancellations arising from unknown physics. The Feynman diagram is depicted in Figure~\ref{fig:mutoeg} and we use the unitary gauge to avoid the need to implement the coupling of $N_{R}$ to the would-be goldstone modes. The amplitude is given by
\begin{equation}
T=A\xi^{\mu}(q)\bar{u}_{e}(p-q)[i\sigma_{\mu\nu}q^{\nu}(1+\gamma^{5})]u_{\mu}(p),
\end{equation}
where $\xi^{\mu}$ denotes the photon polarization vector with momentum $q$. Charged lepton self-energy diagrams and the active neutrino diagrams that are necessary to cancel divergences are not shown. The Majorana neutrino contribution is given by
\begin{equation}
A=\frac{i}{32\pi^{2}}\frac{g^{2}e}{8}\frac{m_{\mu}}{M^{2}_{W}}\epsilon_{\mu}\epsilon_{e}F\left(\frac{M^{2}_{N_{R}}}{M^{2}_{W}}\right),
\end{equation}
where $F(a)$ can be determined from
\begin{equation}
\begin{split}
F(a)&=\frac{13}{6}-\frac{a(1-6a+3a^{2}+2a^{3}-6a^{2}\ln a)}{(1-a)^{4}}\\
    &\equiv \frac{13}{6}-f(a)
\end{split}
\end{equation}
where $a\equiv M^{2}_{N_{R}}/M^{2}_{W}$. The contributions from the active neutrinos can be obtained by taking the limit $M_{N_{R}}\to0$. These contributions can be subtracted out by using the unitarity of the neutrino mixing matrix. With the above two equations, the branching ratio for $\mu\to e\gamma$ is given by
\begin{eqnarray}
BR(\mu\to e\gamma)&=&\frac{3\alpha}{32\pi}|\epsilon_{\mu}\epsilon_{e}|^{2}|f(a)|^2 \nonumber \\
&=&\frac{3\alpha}{8\pi}|\epsilon_{\mu}\epsilon_{e}|^{2},\label{eq:mueg}
\end{eqnarray}
where the second equality is valid in the limit of very heavy $N_{R}$ compare to the gauge boson.

 For the case of right-handed gauge interactions we use the unitarity of the right-handed neutrinos mixing matrix and take the limit where two of the three $N_R^l$
decouple from the low energy theory to obtain
\be
BR(\mu\to e\gamma)=\frac{3\alpha}{32\pi}\left(\frac{M_{W}}{M_{W_{R}}}\right)^{4}|V_{R\,\mu 1} V_{R\,e 1}^*|^{2}|2-f(a)|^2,
\label{eq:muegLR}\ee
where $M_{W}$ denotes the mass of the SM charged gauge boson and $(M_{W}/M_{W_{R}})^{4}$ arises due to normalization. The current experimental bound is $BR(\mu\to e\gamma)\le5\times10^{-13}$~\cite{Renga:2014xra}.

Similarly, the contribution from $N_{R}$ to the anomalous magnetic moment of the muon, $a_{\mu}$ if given by
\begin{eqnarray}
\Delta a_{\mu}&=&-c\left[\frac{5}{12}+\frac{15-74a+114a^{2}-54a^{3}-a^{4}-12a^{2}(2-3a)\ln a}{12(1-a)^{4}}\right] \nonumber \\
&=&-\left\{
\begin{array}{lr}
\frac{5}{3}c,~a\to0 \\
\frac{1}{3}c,~a\to\infty \\
\end{array} \right. \nonumber \\
\label{eq:amu}
\end{eqnarray}
where $c=\frac{g^{2}}{32\pi^{2}}\frac{m^{2}_{\mu}}{M^{2}_{W}}|\epsilon_{\mu}|^{2}$. The first limit in the above equation is consistent with the results appearing in~\cite{Grau:1984zh}. The experimental limit for the deviation is $a^{\text{exp}}_{\mu}-a^{\text{th}}_{\mu}=287(63)(49)\times10^{-11}$. A few comments are in order:
\begin{enumerate}
\item
Neutrinos, heavy or otherwise, contribute negatively to $a_{\mu}$ and will not explain the alleged discrepancy with the SM.
\item
The calculation of $\Delta a_{\mu}$ does not make use of the unitarity condition of the lepton mixing matrix.
\end{enumerate}

Neutrinoless double beta decay can be a sensitive probe for the absolute neutrino mass scale and/or  the mixings of heavy neutrinos with the active ones. The caveat here is that the theoretical uncertainties involved in the calculation of the nuclear matrix elements are difficult to quantify. To compound this  when heavy neutrino intermediate states are
involved one is dealing with very short distances between the two neutrons involved, and the uncertainties are even larger. With this in mind, the best limit on this reaction comes from the decay of $^{76}$Ge which gives an upper bound on the Majorana active neutrino mass of $m_{\nu}<300 - 600$ meV~\cite{KlapdorKleingrothaus:2000sn}. This bound translates to a bound on the light-heavy mixing element $V_{eN}\equiv\epsilon_{e}$, Section~\ref{subsec:onlySM}, and it is given by
\begin{equation}
\frac{|V_{eN}|^{2}}{M_{N_{R}}}<5\times10^{-8}~\text{GeV}^{-1}.
\end{equation}
Within left-right symmetric models, Section~\ref{subsec:LRS}, a similar bound can be obtained on the mixing matrix elements between the right-handed leptons, $V_{ei}$, where $i$ runs over the three heavy right-handed neutrinos, $N^{i=e,\mu,\tau}_{R}$. In particular, one can constrain
\begin{equation}
\epsilon_{N_{R}}=\sum^{3}_{i=1}V^{2}_{R\,ei}\frac{m_{p}}{M_{N_{R}}}\frac{M^{4}_{W}}{M^{4}_{W_{R}}},
\end{equation}\cite{Das:2012ii}
using the current experimental limit of $|\epsilon_{N}|<2\times10^{-8}$~\cite{Rodejohann:2011mu}.

In addition to low energy direct and indirect constraints, collider searches can be used to probe the single Majorana extension of the SM, in particular flavor violating final states. The most recent analysis by the CMS collaboration targets the same-sign dimuon channel with $\sqrt{s}=8$ TeV center of mass energies and $19.7$ fb$^{-1}$ of integrated luminosity~\cite{Khachatryan:2015gha}. The limits obtained are on the $V_{\mu N}$ mixing element and result on the following upper bounds: $|V_{\mu N}|^{2}<0.00470,0.0123,0.583$ for Majorana neutrinos masses of $90,200$ and $500$ GeV respectively. However, for a  $200$ and $300$ GeV Majorana mass the unitarity bound given in Eq.~(\ref{eq:lepU}) is stronger, making the latter the primary constraint limiting the value of $\epsilon_{\mu}$ that we use in our analysis. This analysis equally constrains the left-right symmetric model discussed in the previous section. In addition, both CMS and ATLAS have reported limits on heavy Majorana neutrino production in the context of the left-right symmetric models~\cite{Khachatryan:2014dka,ATLAS:2012ak}. The limits are for heavy Majorana production with $M_{W_{R}}>M_{N_{R}}$ and identical left- and right-handed mixing matrices. The region excluded extends to $M_{W_{R}}\sim 3$ TeV, with a region defined in the $M_{W_{R}}-M_{N_{R}}$ plane.

\section{Same-sign leptons at the LHC and beyond}\label{sec:Signal}

\subsection{Background}\label{subsec:Bkg}
Within the SM, lepton number violating processes are absent at zero temperature. However, there exist certain SM processes that can give rise to a same-sign lepton final state in association with jets. The two leading backgrounds discussed below are simulated at leading order using MadGraph 5~\cite{Alwall:2011uj}. We implement PYTHIA~\cite{Sjostrand:2006za} for the parton showering and hadronization and a fast detector simulation is carried out using Delphes 3~\cite{delphes}. The detector simulator is used for jet clustering, track reconstruction and lepton identification and isolation. Light jets are reconstructed using the anti-$k_{T}$ algorithm with parameter $\Delta R=0.5$. Furthermore, we calculate the lepton isolation criteria, $I_{l}$, by summing the $p_{T}$ of all tracks within a cone of size $\Delta R=0.5$ around the leptons and require that
\begin{equation}
 I_{l}=\frac{\sum_{i\ne l}p_{T,i}}{p_{T,l}} < 0.10.
 \end{equation}
The two leading background processes at the LHC are the following:
\begin{enumerate}
\item
{\bf $p~p\to W^{+}W^{+}j~j$ with low missing transverse energy, $\slashed{E}_{T}$.} \newline
\\
This SM process is an irreducible background and dominates when the amount of missing transverse energy is small and comparable to the signal. The parameter space for heavy Majorana neutrinos within the SM was studied in~\cite{Han:2006ip} using $14$ TeV center of mass energies and a luminosity of $100$ fb$^{-1}$. The authors identified the following series of cuts that would suppress this background and enhance the significance of a  $\mu^{\pm}\mu^{\pm}j~j$:
\begin{eqnarray}
p_{T,\mu}&>&10~\text{GeV},~~~~~ \eta_{\mu}<2.5 \nonumber \\
p_{T,j}&>&15~\text{GeV},~~~~~ \eta_{j}<3.0 \nonumber \\
\Delta R^{\text{min}}_{lj}&>&0.5,~~~~~~\slashed{E}_{T}<25~\text{GeV}.
\end{eqnarray}
However, a search performed by the CMS collaboration for same-sign muons and electrons using $4.9$ fb$^{-1}$ of data at $\sqrt{s}=7$ TeV~\cite{Giordano:2013xba} determined that the leading background for same-sign leptons in association with jets is from the QCD multijet component with misidentify jets as leptons. Consequently, the impact of the QCD background should be analyzed further.
\item
{\bf QCD multijet background.} \newline
\\
This background is mostly due to $p~p\to 4j$ with two jets misidentified as leptons. In addition, leptons may arise from decays of heavy flavored jets. Although there are studies by ATLAS on muon fake rates~\cite{ATLASfakemu}, the electron fake rate does not have a reference value. However, from the studies in~\cite{Giordano:2013xba} and~\cite{ATLASfakemu}, the electron fake rate can be anywhere between $10^{-4}-10^{-5}$. In what follows we refer to the fake rate as $\epsilon_{j\to l}$ and emphasize that the fake rate highly depends on the detector and energy, making it difficult to determine with a fast detector simulator such as Delphes. In our analysis, we will assume a given fake rate and determine the background suppression as a function of the misidentification efficiency. We extract an upper bound on this rate by comparing our simulated background after applying the following kinematic cuts that we find to significantly enhance our signal over the irreducible background described above:
 \begin{eqnarray}
p_{T,\mu}&>&10~\text{GeV},~~~~~ \eta_{\mu}<2.5 \nonumber \\
p_{T,j}&>&20~\text{GeV},~~~~~ \eta_{j}<5.0 \nonumber \\
m_{l_{1},l_{2}}&>&10~\text{GeV},~~~~\slashed{E}_{T}<40~\text{GeV},\label{eq:Dcut}
\end{eqnarray}
where $l_{1}$ and $l_{2}$ denote the leading an subleading leptons, to the number of simulated events using the latest search for $\mu^{\pm}\mu^{\pm}$ by the CMS collaboration~\cite{Giordano:2013xba} with $19.7$ fb$^{-1}$ at $\sqrt{s}=8$ TeV. The result is a muon misidentification rate (fake rate) of $\epsilon_{j\to l}\sim 2\times10^{-5}$. We use a generous upper bound for the electron misidentification rate of $10^{-5}$.
\end{enumerate}
\begin{table}[ht]\centering
 \tabcolsep 2.2 pt
\small
\begin{tabular}{|c|c|c|c|c|c|}
\hline
     &  & $e^{+}e^{+}\nu\nu~jj$ & $e^{+}\mu^{+}\nu\nu~jj$ & $\mu^{+}\mu^{+}\nu\nu~jj$ & $4j\cdot\epsilon^{2}_{j\to l}$  \\
\hline
\hline
Parton level             &      $\sigma$ (fb)       &	0.3927		&	0.7849	&	0.3927	& 1.79 \\
\hline
            &      $N$       &	1178		&	2354 	&	1178 	&  5370 \\
\hline
Detector level +~(\ref{eq:Dcut})            &        $\sigma$ (fb)      &		0.1187	&	0.2674	&	0.1187	&  0.471\\
\hline
           &         $N$    &	356		&	802	&	356	&  1410\\
\hline
\parbox[t]{0.40\textwidth}{ \centering Detector level +~(\ref{eq:Dcut}) \\ +$(M_{W} -10.) < m_{j_{1},j_{2}}<(M_{W} +10.)$  }            &    $\sigma$ (fb)         &		 0.0094	 &	0.0202	&	0.0119	& 0.0643 \\
\hline
            &      $N$       &	28.3		&	60.7	&	35.6	&  193\\
\hline
\parbox[t]{0.40\textwidth}{ \centering Detector level +~(\ref{eq:Dcut}) \\ +$(M_{W} -5.) < m_{j_{1},j_{2}}<(M_{W} +5.)$  }               &      $\sigma$ (fb)        &		 0.0047	 &	0.0100	&	0.0059	& 0.0323\\
\hline
            &      $N$       &	14.18		&	30	&	17.6	&  96.9 \\
\hline
\end{tabular}
\caption{\small SM backgrounds at $13$ TeV. The event number is shown at a luminosity of $3000$ fb$^{-1}$.} \label{tab:13TEVbkg}
\end{table}
\begin{table}[ht]\centering
 \tabcolsep 2.2 pt
\small
\begin{tabular}{|c|c|c|c|c|c|}
\hline
     &  & $e^{+}e^{+}\nu\nu~jj$ & $e^{+}\mu^{+}\nu\nu~jj$ & $\mu^{+}\mu^{+}\nu\nu~jj$ & $4j\cdot\epsilon^{2}_{j\to l}$  \\
\hline
\hline
Parton level             &      $\sigma$ (fb)       &	3.686		&	7.370	&	3.694	& 44.2\\
\hline
            &      $N$       &	11058		&	22110	&	11082	& 133000\\
\hline
Detector level +~(\ref{eq:Dcut})             &        $\sigma$ (fb)      &		0.962	&	2.148	&	1.189	& 11.6\\
\hline
           &         $N$    &	2887		&	6444 	&	3568 	& 35000\\
\hline
\parbox[t]{0.40\textwidth}{ \centering Detector level +~(\ref{eq:Dcut}) \\ +$(M_{W} -10.) < m_{j_{1},j_{2}}<(M_{W} +10.)$  }            &    $\sigma$ (fb)         &		 0.0597	 &	0.1388	&	0.0784	& 1.59\\
\hline
            &      $N$       &	179		&	416	&	235	& 4770\\
\hline
\parbox[t]{0.40\textwidth}{ \centering Detector level +~(\ref{eq:Dcut}) \\ +$(M_{W} -5.) < m_{j_{1},j_{2}}<(M_{W} +5.)$  }               &      $\sigma$ (fb)        &		 0.0291	 &	0.0692	&	0.0395	& 0.798\\
\hline
            &      $N$       &	87.2		&	207.6	&	118.4	& 2390\\
\hline
\end{tabular}
\caption{\small SM backgrounds at $100$ TeV. The event number is shown at a luminosity of $3000$ fb$^{-1}$.} \label{tab:100TEVbkg}
\end{table}
In Tables~\ref{tab:13TEVbkg} and~\ref{tab:100TEVbkg} we show the generated backgrounds at parton level and after a full fast detector simulation for $\sqrt{s}=13,100$ TeV center of mass energies respectively with $3000$ fb$^{-1}$ of integrated luminosity. The detector level results are shown applying the cuts discussed in Equation~(\ref{eq:Dcut}) and the requirement that the leading two jets arise from the decay of a SM $W$ gauge boson. The latter is used to enhance the sensitivity to the single heavy Majorana neutrino extension to the SM discussed in Section~\ref{subsec:onlySM}.

\subsection{SM + $N_{R}$: S-channel collider reach at $13$ and $100$ TeV}\label{subsec:Schannel}
\begin{figure}[ht]\centering
\begin{subfigure}[b]{0.48\textwidth}
\includegraphics[width=\textwidth]{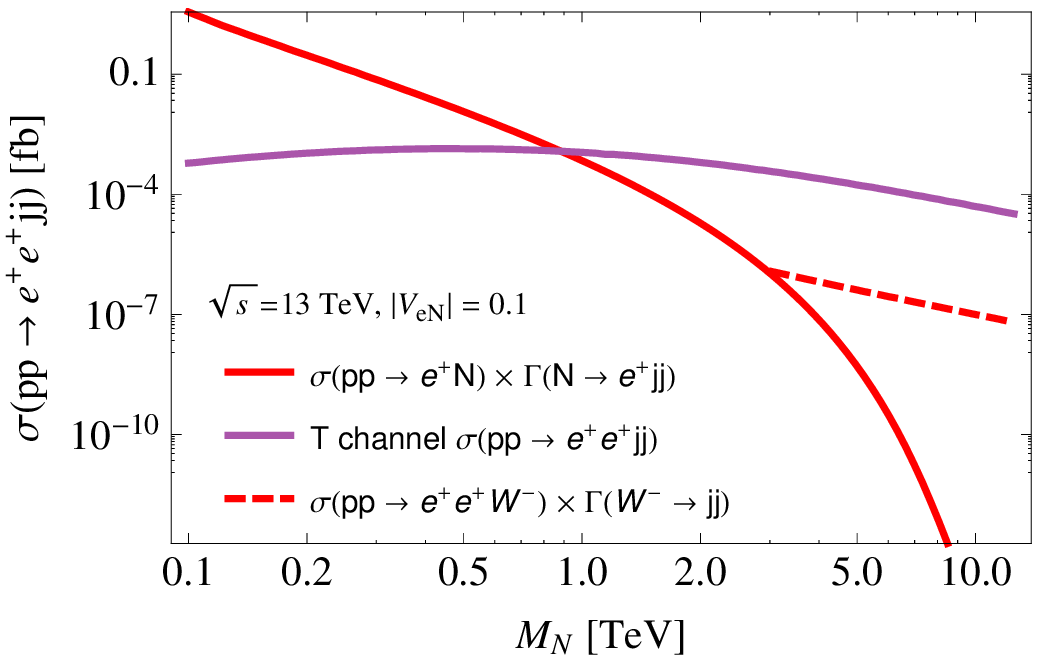}\subcaption{}
\end{subfigure}
\begin{subfigure}[b]{0.48\textwidth}
\includegraphics[width=\textwidth]{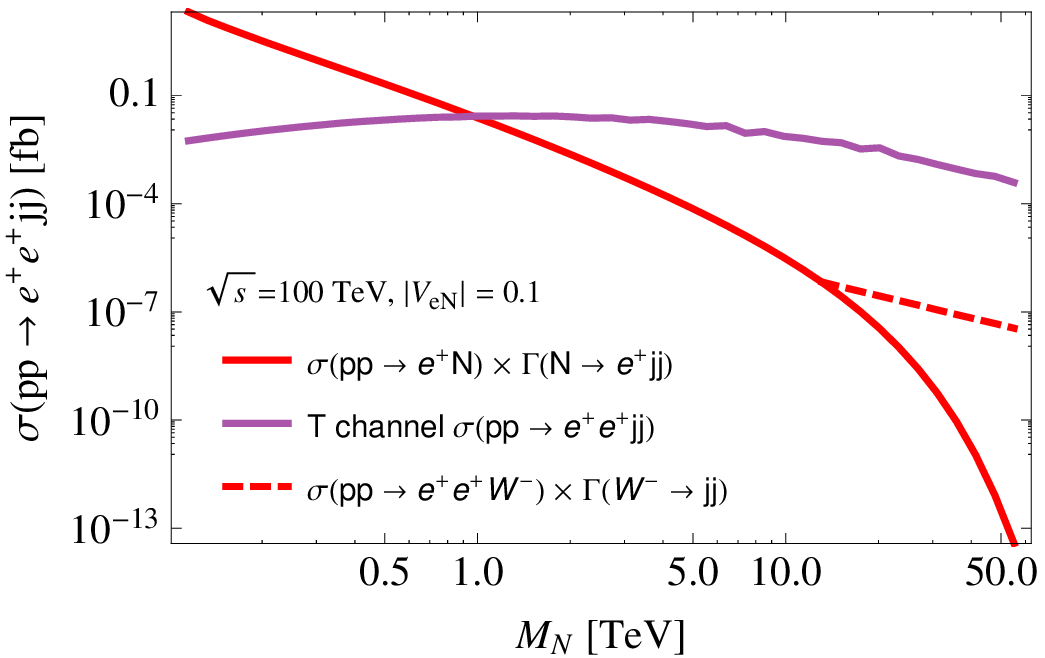}\subcaption{}
\end{subfigure}
\caption{\small The $S$- and $T$-channel contribution using default parton-level cuts implemented in MadGaph 5 at $13$ (left) and $100$ (right) TeV using a flavor universal mixing angle of $\epsilon=0.1$.}\label{fig:SvsT}
\end{figure}
The presence of heavy Majorana neutrinos in either a structure only consisting of SM gauge symmetries or an extension of the SM with an additional non-abelian group structure will contribute to the process $p~p\to l^{\pm}l^{\pm}~jj$ in two topological classes: $S$- and $T$-channels. The $S$-channel topology is depicted in Figure~\ref{fig:schannel} and it is dominated by quark anti-quark fusion, in particular, the first generation. The $T$-channel topology is depicted in Figure~\ref{fig:WWfusion} and has contributions from both quark-quark and quark-antiquark fusion. In the massless limit, for quarks, these two contributions have different helicity amplitudes. In this section we examine the sensitivity that Run II at the LHC and a future 100 TeV collider will have to the model discussed in Sections~\ref{subsec:onlySM}. For the left-right symmetric model, a combined $S$- and $T$-channel analysis is performed, since within this theoretical framework, the mass of the SU(2)$_{\text{R}}$ gauge boson is also a free parameter. The simulation of the signal at the parton level is carried out using MadGraph 5~\cite{Alwall:2011uj}. In Figures~\ref{fig:SvsT}(a) and~\ref{fig:SvsT}(b) we show the $S$- and $T$- channel contributions to the same-sign electron cross section for the LHC running at $\sqrt{s}=14$ TeV (left) and $\sqrt{s}=100$ TeV (right) within the framework a single heavy Majorana neutrino extension of the SM using a coupling of $V_{eN}=0.1$. From the plots we can observe that the $T$-channel contribution to the production cross section starts dominating for masses, $M_{N_{R}}$, approximately above $1$ TeV.

In the above section we introduced the two most dominant backgrounds that must be suppressed to enhance the sensitivity that a hadron collider requires to probe a model with a lepton number violating same-sign lepton final state. Below we examine and introduce a search strategy aimed at extracting a statistical significant signal for Majorana masses below $1$ TeV. The constraints discussed in Section~\ref{subsec:constraints}, in particular $\mu\to e \gamma$, narrow down the relevant final states to $e^{\pm}e^{\pm}j~j$ or $\mu^{\pm}\mu^{\pm}j~j$, since in either case the cross section depends only on $\epsilon_{e}$ or $\epsilon_{\mu}$, and we can use either coupling to suppress the $\mu\to e\gamma$ constraint.

The simulation of the signal at the parton level is carried out using MadGraph 5~\cite{Alwall:2011uj} with model files generated with FeynRules~\cite{Alloul:2013bka}.
We use two input parameters in our simulation, the mass of the heavy Majorana neutrino, $M_{N_{R}}$, and a universal mixing to the active neutrino species, $\epsilon_{e}=\epsilon_{\mu}=\epsilon=0.1$. We separately generate $e^{\pm}e^{\pm}j~j$ and $\mu^{\pm}\mu^{\pm}j~j$ final states. In addition we focus on Majorana masses above $100$ GeV to avoid constraints from LEP~\cite{Achard:2001qv,Abreu:1996pa,Akrawy:1990zq} that restrict $|\epsilon_{\mu}|^{2}$ to values below $10^{-4}-10^{-5}$ for Majorana masses between $5-80$ GeV. The partial widths of $N_{R}$ are thus given by
\begin{equation}
\Gamma_{N\to l^{+}W^{-}}\approx\Gamma_{N\to l^{-}W^{+}}\approx \Gamma_{N\to \nu Z}\approx \frac{g^{2}\epsilon^{2}}{64\pi}M^{2}_{N_{R}}\left[\frac{M^{2}_{N_{R}}}{M^{2}_{W}}+1\right].
\end{equation}

\begin{figure}[ht]\centering
\begin{subfigure}{0.48\textwidth}
\includegraphics[width=\textwidth]{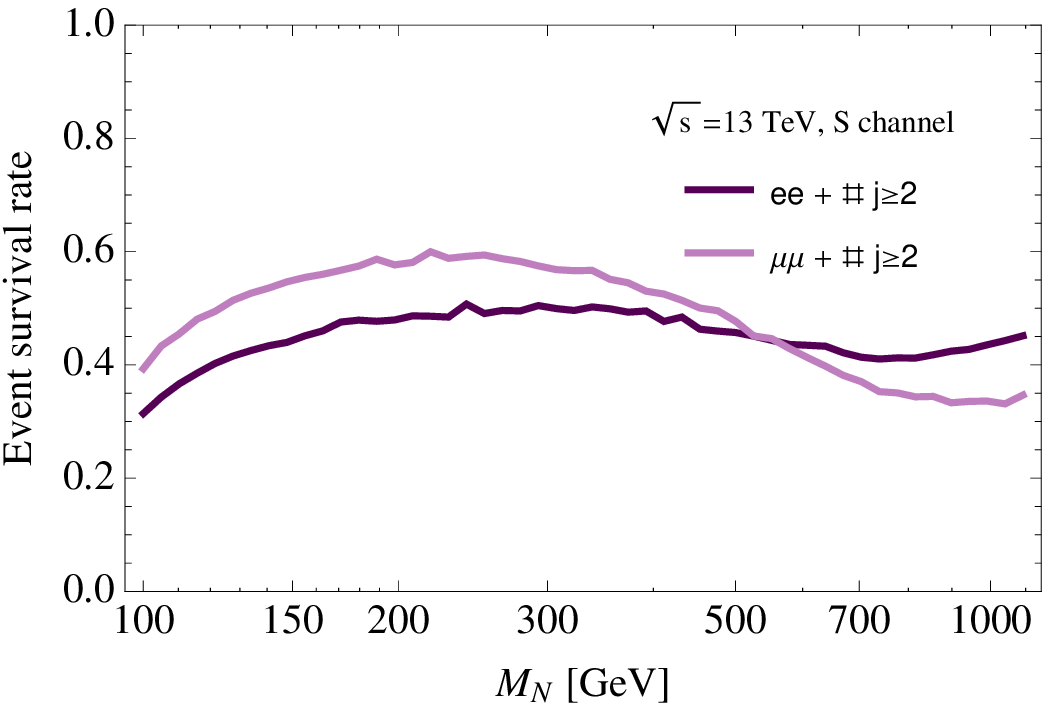}\subcaption{}
\end{subfigure}
\begin{subfigure}{0.48\textwidth}
\includegraphics[width=\textwidth]{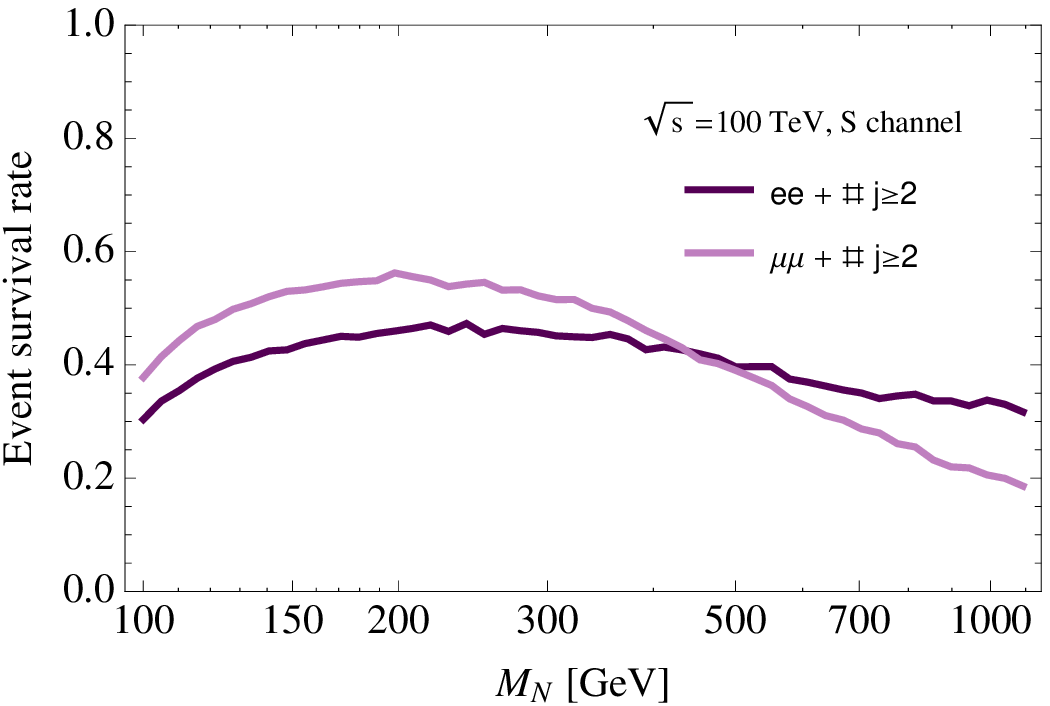}\subcaption{}
\end{subfigure}
\caption{\small Survival rate for a same-sign lepton final state with at least two jets for $13$ TeV (left) and $100$ TeV (right) center of mass energies.}\label{fig:SsurviveS}
\end{figure}
\begin{figure}[ht]
\begin{center}
\includegraphics[width=3.0in]{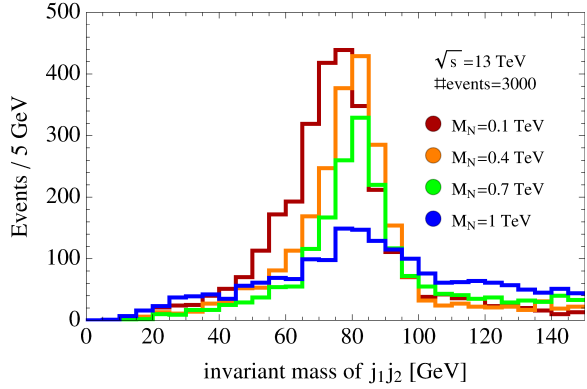}
\caption{\small Invariant mass distribution of the two leading jets in our $eejj$ signal for various heavy Majorana neutrino masses, $M_{N_{R}}$.}\label{fig:Wpole}
\end{center}
\end{figure}

We begin the analysis by applying the cuts used to treat the backgrounds described in Equation~(\ref{eq:Dcut}). However, the full detector simulation misidentifies a number of leptons as jets. The survival rate for our signal is shown in Figure~\ref{fig:SsurviveS}(a) for a $13$ TeV collider and in Figure~\ref{fig:SsurviveS}(b) for $100$ TeV. A clear signature of this model is the $W$ that arises from the decay of a Majorana neutrino. Since we are looking at the hadronic decay mode of the gauge boson, the $W$ can be tagged using the invariant mass of two jets in the event, in particular the leading two jets. In Figure~\ref{fig:Wpole} we show the invariant mass for different masses of the Majorana neutrino, $M_{N_{R}}$.

To analyze the reach at the LHC and in a future $100$ TeV collider to the S-channel signal of the simplified model, we define a significance variable by
\begin{equation}
\chi=\frac{S}{\sqrt{S+B}}
\end{equation}
where $S$ denotes the number of signal events and $B$ the number of background events. Since we are using an inclusive final state (in charge) with same-sign leptons, the production cross-section is given by
\begin{eqnarray}
\sigma&\propto&\epsilon^{2}\left[\Gamma(u,\bar{d}\to W^{*})BR(N\to l^{+}+jj')+\Gamma(u,\bar{d}\to W^{*})BR(N\to l^{-}+jj')\right] \nonumber \\
&=&\epsilon^{2}\left[\Gamma(u,\bar{d}\to W^{*})+\Gamma(u,\bar{d}\to W^{*})\right]BR(N\to l^{\pm}+jj').
\end{eqnarray}
Given that $N_{R}$ is Majorana, $BR(N_{R}\to l^{+}+jj')=BR(N_{R}\to l^{-} jj')$ and given the fact that they are independent of the momentum transfer, we can treat $\epsilon^{2}BR(N\to e^{\pm}+jj')$ as a scaling factor to the total amplitude. Therefore, we choose to carry out the simulation for a fixed value of $\epsilon=0.1$ for all three lepton flavors. Thus
\begin{equation}
\epsilon^{2} BR(N\to e^{\pm}+jj')_{\epsilon=V_{e4}=V_{\mu 4}=V_{\tau 4}}=|V_{e4}|^{2}BR(N\to e^{\pm}+jj'),
\end{equation}
and since $BR(N\to e^{\pm}+jj')=2/27$ for the range of masses that we are considering, we can write
\begin{equation}
\frac{2}{27}\epsilon^{2}=|V_{e4}|^{2}BR(N\to e^{\pm}+jj').
\end{equation}
Similarly for the $\mu\mu jj$ final states. In our analysis we include systematic and statistical uncertainties and model the distribution as Gaussian with an statistical error given by $\sqrt{N}$, for $N$ number of events. We find the $95\%$ confidence level fit to $\epsilon$ using a $\chi$ variable defined by
\begin{equation}
\chi=\frac{\hat{\epsilon}^{2}S_{0}}{\sqrt{\hat{\epsilon}^{2}S_{0}+B_{FR}+B_{SM}}+\sigma_{sys}},\label{eq:chiVar}
\end{equation}
\begin{figure}[ht]\centering
\begin{subfigure}{0.48\textwidth}
\includegraphics[width=\textwidth]{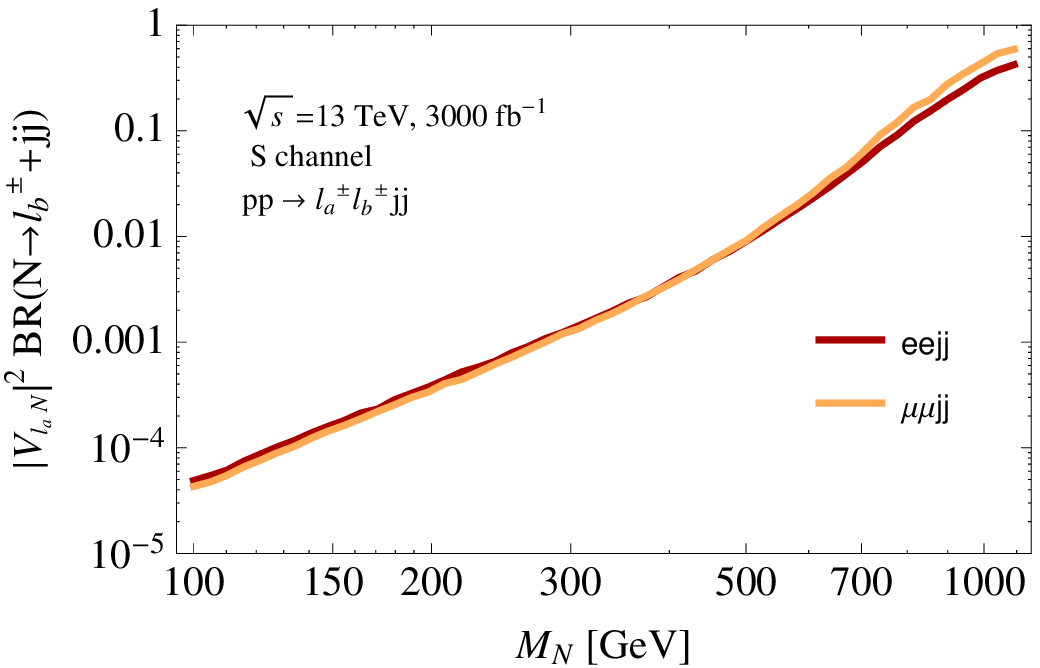}\subcaption{}
\end{subfigure}
\begin{subfigure}{0.48\textwidth}
\includegraphics[width=\textwidth]{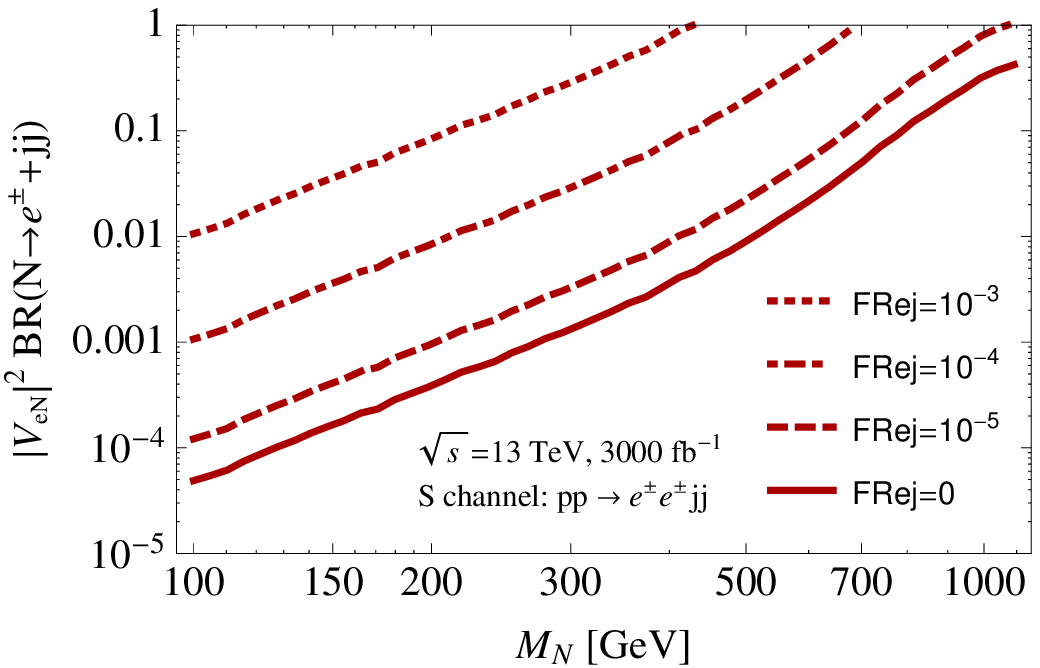}\subcaption{}
\end{subfigure}
\caption{\small The $95\%$ expected limit at $13$ TeV and $3000$ fb$^{-1}$ of integrated luminosity. On the left, the limits are shown without the QCD contribution while on the right, varying fake rates are introduced for the $e^{\pm}e^{\pm}jj$ final state. }\label{fig:fig7}
\end{figure}
where $\hat{\epsilon}=\epsilon/0.1$, $S_{0}$ denotes the number of signal events with the benchmark value $\epsilon=0.1$ and $B_{SM},B_{FR}$ denote the SM irreducible and QCD backgrounds respectively. The latter is scaled with the fake rate described earlier in the section. The systematic uncertainty is parametrized by $\sigma_{sys}$ and we assume it to be negligible compared to the statistical component. In Figures~\ref{fig:fig7}(a) we plot the $95\%$ C.L region of parameter space against the irreducible background using a center of mass energy of $13$ TeV and $3000$ fb$^{-1}$ of data. In Figure~\ref{fig:fig7}(b) we show the effect the QCD background with different fake rates in the $eejj$ channel. In Figures~\ref{fig:fig8}(a) and~\ref{fig:fig8}(b) we show the results of the simulation for a $100$ TeV collider with $3000$ fb$^{-1}$ of data.
\begin{figure}[ht]\centering
\begin{subfigure}{0.48\textwidth}
\includegraphics[width=\textwidth]{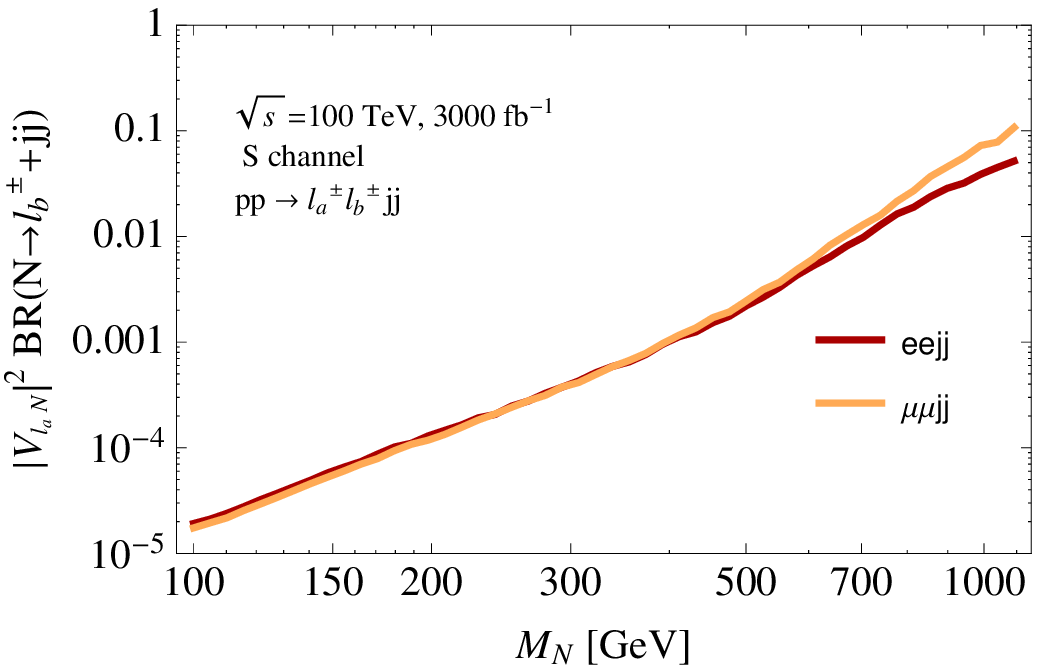}\subcaption{}
\end{subfigure}
\begin{subfigure}{0.48\textwidth}
\includegraphics[width=\textwidth]{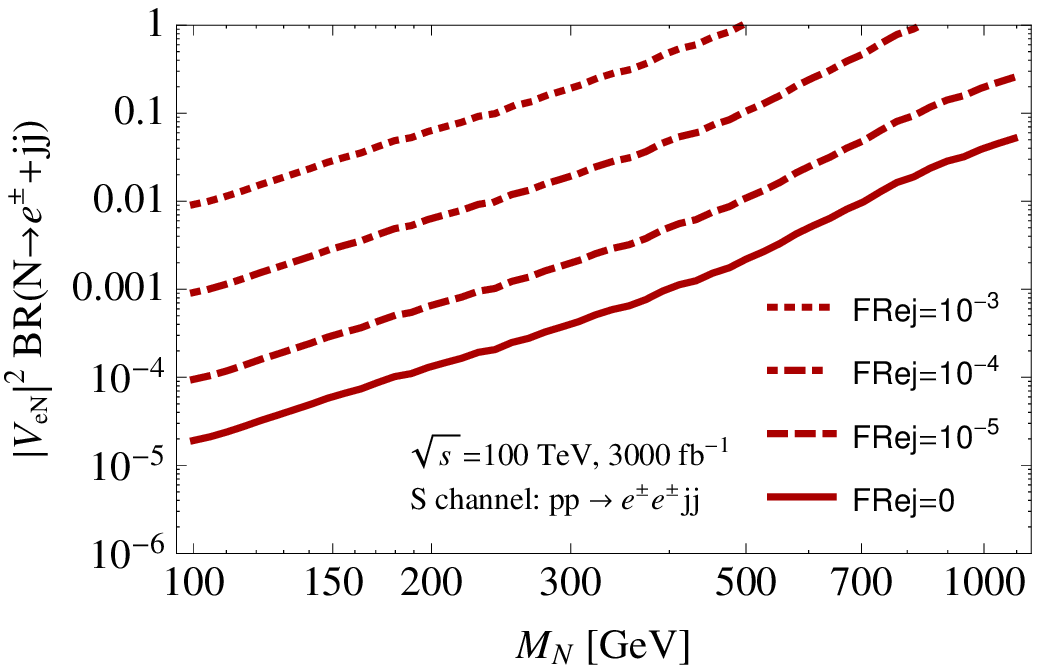}\subcaption{}
\end{subfigure}
\caption{\small The $95\%$ expected limit at $100$ TeV and $3000$ fb$^{-1}$ of integrated luminosity. On the left, the limits are shown without the QCD contribution while on the right, varying fake rates are introduced for the $e^{\pm}e^{\pm}jj$ final state. }\label{fig:fig8}
\end{figure}
It is worth mentioning that our results improve by a non-negligible amount if instead of doing our fit to the the whole range of Majorana masses, we carry out a fit to bins of the Majorana mass parameter, $M_{N_{R}}$. In particular, we define a bin size of $10$ GeV and fit our simulated data to obtain the $95\%$ C.L regions; these are depicted in Figures~\ref{fig:fig9}(a) and~\ref{fig:fig9}(b) together with the results shown in the previous two figures for comparison. It is worth pointing out then that our analysis shows that a $13$ TeV machine will have sensitivity to couplings approximately an order of magnitude below those probed by the latest CMS same-sign muon search~\cite{Khachatryan:2015gha}.
\begin{figure}[ht]\centering
\begin{subfigure}{0.48\textwidth}
\includegraphics[width=\textwidth]{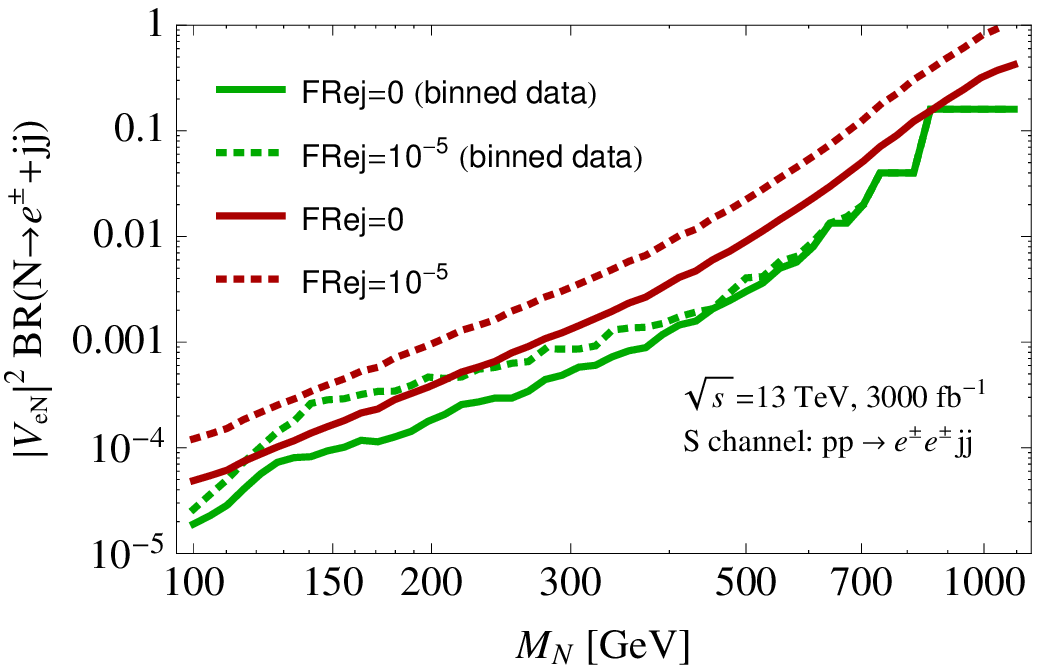}\subcaption{}
\end{subfigure}
\begin{subfigure}{0.48\textwidth}
\includegraphics[width=\textwidth]{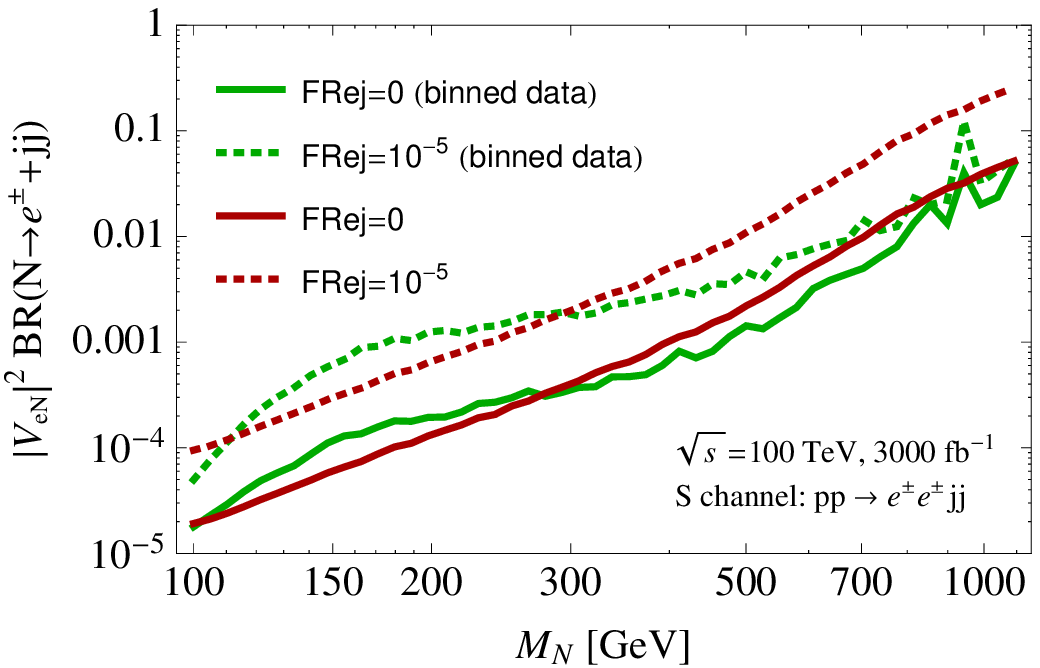}\subcaption{}
\end{subfigure}
\caption{\small The $95\%$ expected limits with a fit to $10$ GeV bins of the invariant mass, $M_{ljj}$. The plot on the left corresponds to $13$ TeV center of mass energies with varying fake rates while the plot on the right is a $100$ TeV machine. For comparison the unbinned fit is shown in red.}\label{fig:fig9}
\end{figure}

\subsection{SM + $N_{R}$: T-channel collider reach at $13$ and $100$ TeV}\label{subsec:Tchannel}

We saw at the beginning of Section~\ref{subsec:Schannel} that the $S$-channel contribution to the production cross section decreases significantly for Majorana masses above $\sim1$ TeV and the $T$-channel contribution begins to dominate. This effect can be easily understood given that for quark-antiquark invariant masses below the mass of the heavy neutrino, the $pp\to l^{+}~N_{R}$ is closed while the $pp\to l^{+}l^{+}~W^{-}$ is open; that is
\begin{eqnarray}
s &>& M^{2}_{N_{R}},~~\sigma\left(pp\to l^{+}l^{+}~jj\right)_{\text{dominant}}=\sigma\left(pp\to l^{+}~N_{R}\right)BR(N_{R}\to l^{+}~jj) \nonumber \\
s &<& M^{2}_{N_{R}},~~\sigma\left(pp\to l^{+}l^{+}~jj\right)_{\text{dominant}}=\sigma\left(pp\to l^{+}l^{+}~W^{-}\right)BR(W\to jj).
\end{eqnarray}
Although the phase space contribution from a $3$-particle final state is larger than a $4$-particle final state ($T$-channel), the $S$-channel amplitude is greatly affected due to the PDF suppression at high momentum fraction of the quark-antiquark initial state. A naive estimate allows us to determine that a $13$ TeV $pp$ machine with $3000$ fb$^{-1}$ of data cannot extract a statistically significant signal for masses, $M_{N_{R}}$, in the range $1-10$ TeV. However, a $100$ TeV machine with $3000$ fb$^{-1}$ of integrated luminosity can yield approximately $30$ events with $\epsilon=0.1$ before a full detector simulation is implemented. In Appendix~\ref{app:AppA} we show an explicit analysis of the various contributions to the $T$-channel amplitude that is used to determine a series of efficient cuts to extract a signal for Majorana masses above $1$ TeV. In particular, we show that the amplitudes are proportional to the jet-pair invariant mass with the strongest enhancement arising from contributions proportional to the Majorana mass. In addition, for momentum transfer, $|q|^{2}$, below $M_{N_{R}}$, the amplitudes are also proportional to the lepton pair invariant mass.


\begin{figure}[ht]\centering
\begin{subfigure}{0.48\textwidth}
\includegraphics[width=\textwidth]{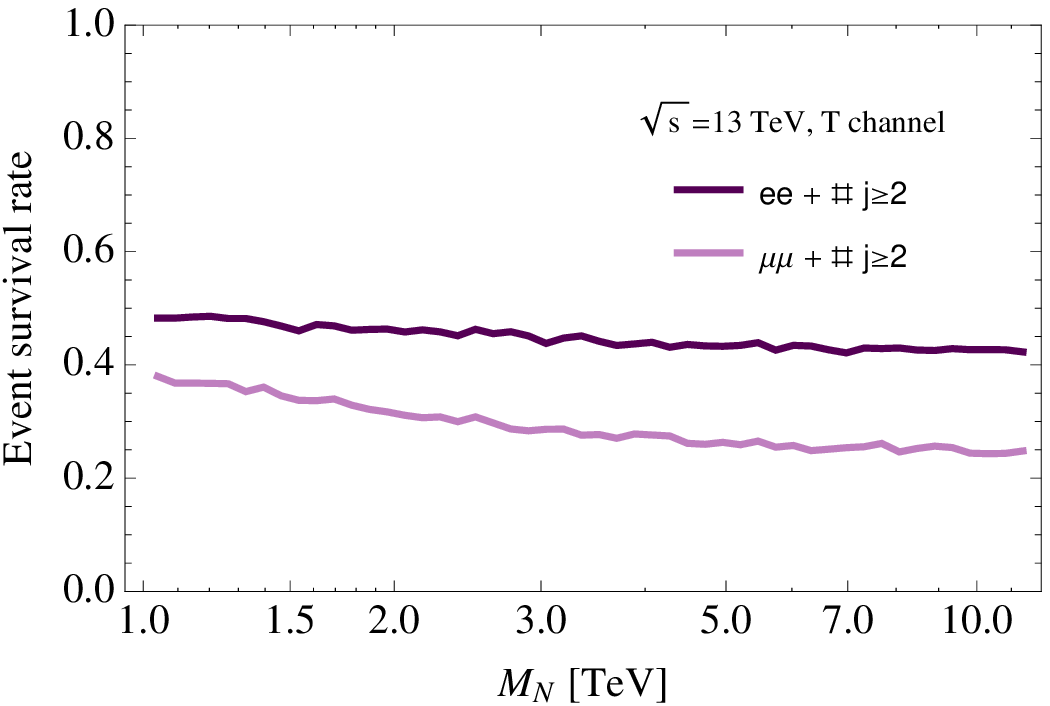}\subcaption{}
\end{subfigure}
\begin{subfigure}{0.48\textwidth}
\includegraphics[width=\textwidth]{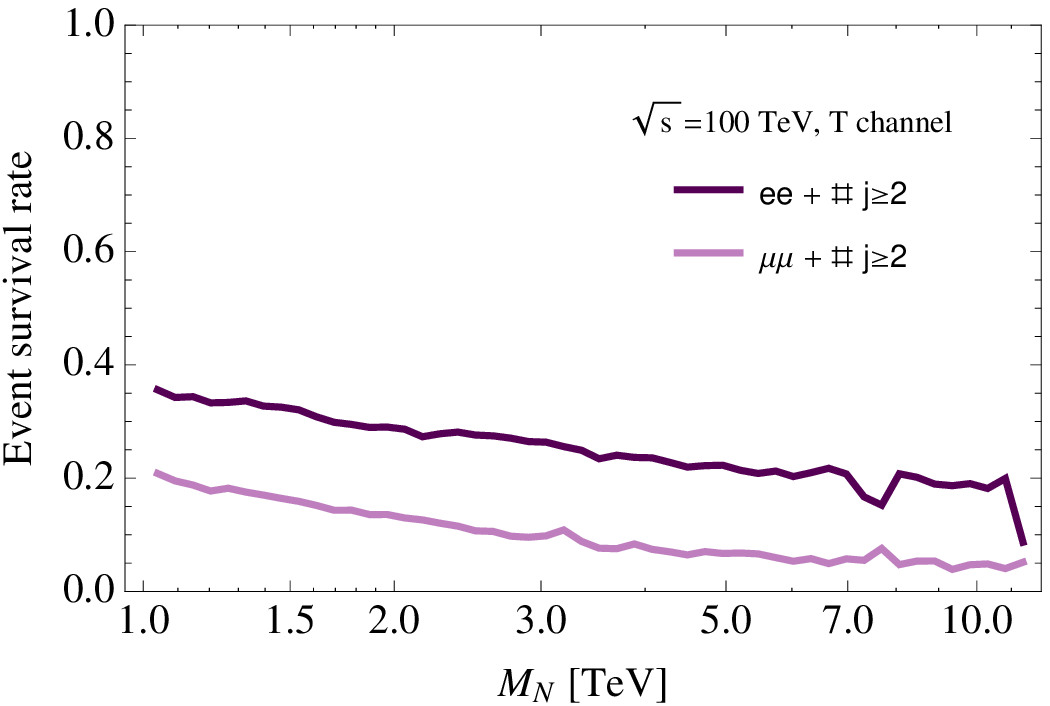}\subcaption{}
\end{subfigure}
\caption{\small Survival rate for a same-sign lepton final state with at least two jets for $13$ and $100$ TeV center of mass energies.}\label{fig:Fig10}
\end{figure}
\begin{figure}[ht]\centering
\begin{subfigure}{0.48\textwidth}
\includegraphics[width=\textwidth]{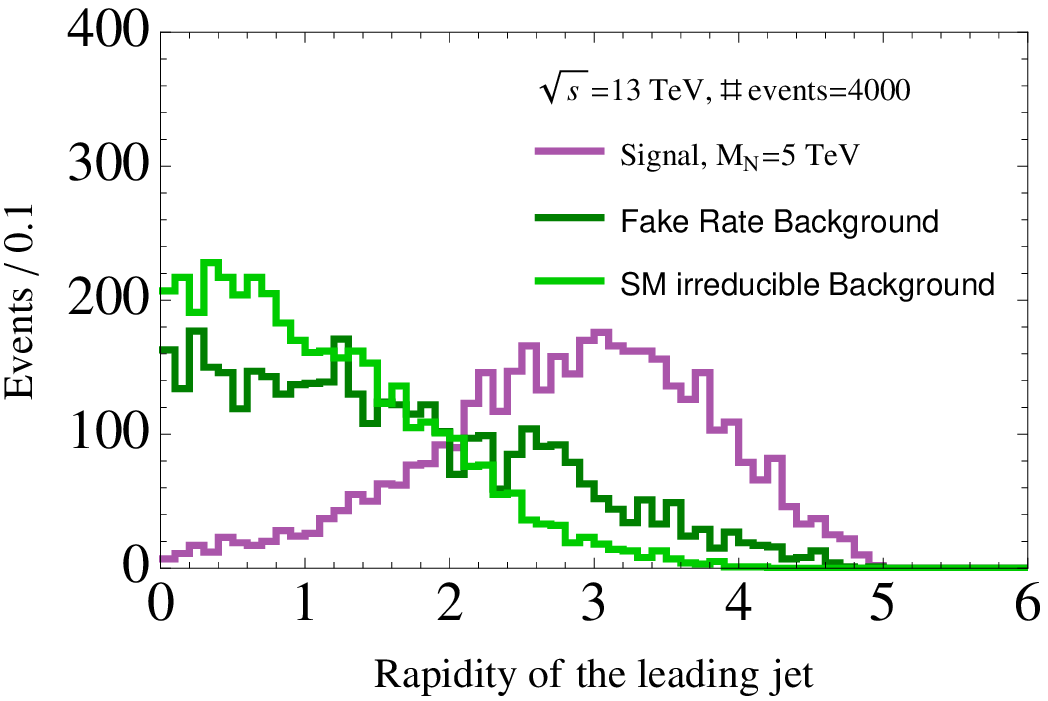}\subcaption{}
\end{subfigure}
\begin{subfigure}{0.48\textwidth}
\includegraphics[width=\textwidth]{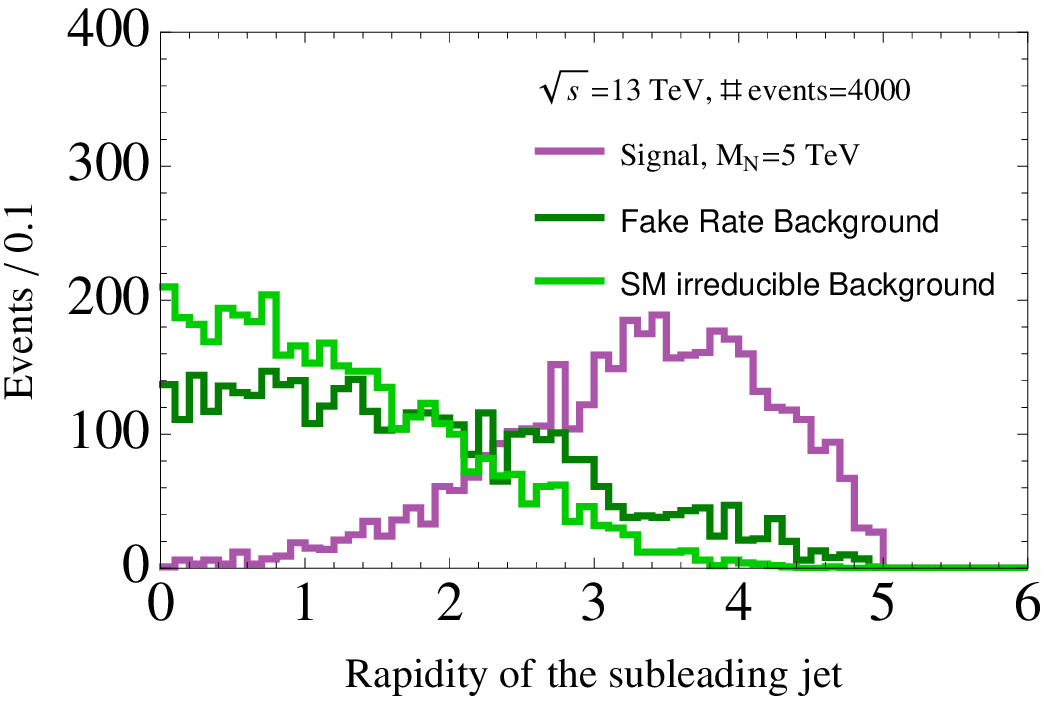}\subcaption{}
\end{subfigure}
\caption{\small Pseudorapidity distribution of leading (left) and subleading (right) jets for the signal and SM backgrounds using a heavy Majorana neutrino with mass $M_{N_{R}}=5$ TeV.}\label{fig:Fig11}
\end{figure}

In Figure~\ref{fig:Fig10} we show the performance of the detector simulation after implementing the default cuts introduced in Equation~(\ref{eq:Dcut}). For the range of masses that we are interested in we show the same-sign lepton reconstruction efficiency as a function of $M_{N_{R}}$ in both $13$ and $100$ TeV machines. In both cases, the overall drop in efficiency is mainly due to the large pseudorapidities associated with jets in the hard process since the track reconstruction efficiency drops significantly for values of $|\eta|$ above $2.5$. In fact, from Figure~\ref{fig:Fig11} we can see how boosted jets are in comparison to the SM background.

\begin{figure}[ht]\centering
\begin{subfigure}{0.48\textwidth}
\includegraphics[width=\textwidth]{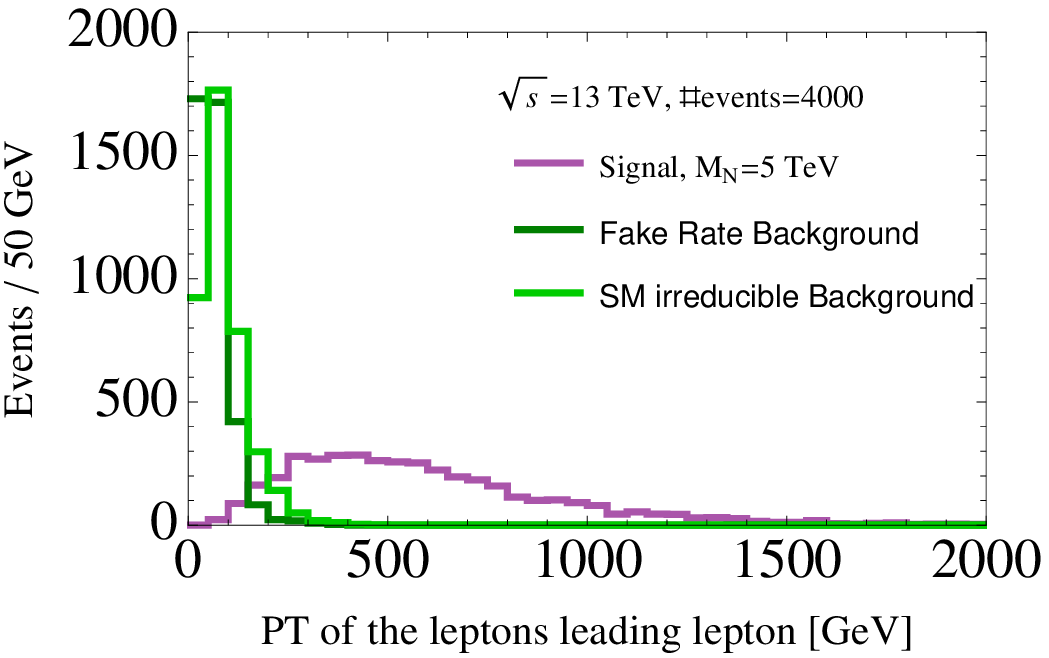}\subcaption{}
\end{subfigure}
\begin{subfigure}{0.48\textwidth}
\includegraphics[width=\textwidth]{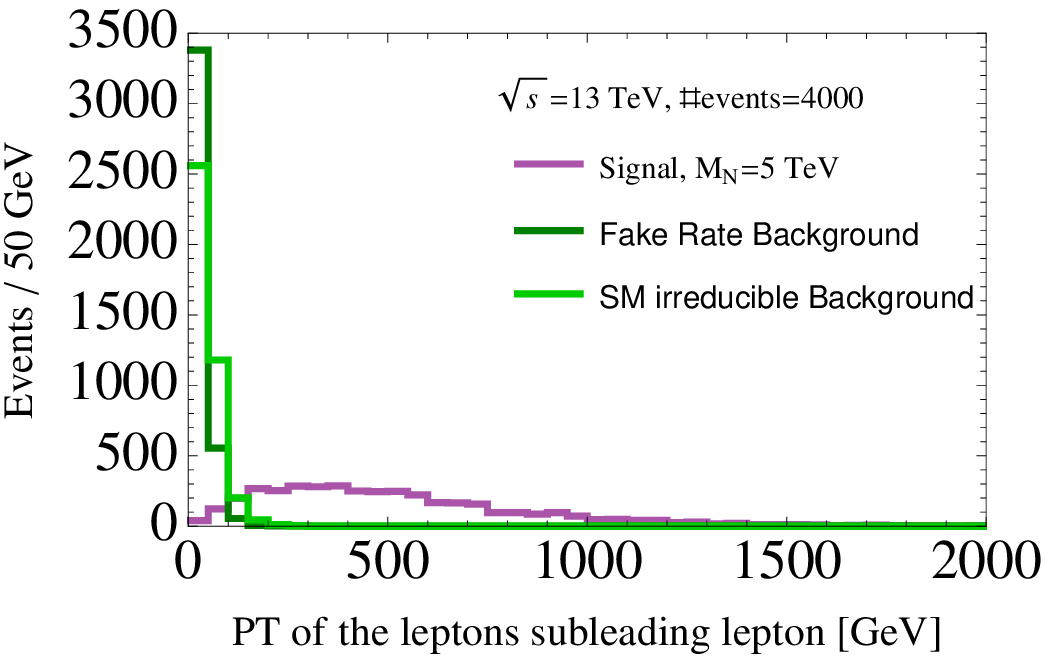}\subcaption{}
\end{subfigure}
\caption{\small $p_{T}$ distribution of leading (left) and subheading (right) leptons for the signal and SM backgrounds using a heavy Majorana neutrino with mass $M_{N_{R}}=5$ TeV.}\label{fig:Fig12}
\end{figure}

\begin{figure}[ht]\centering
\begin{subfigure}{0.48\textwidth}
\includegraphics[width=\textwidth]{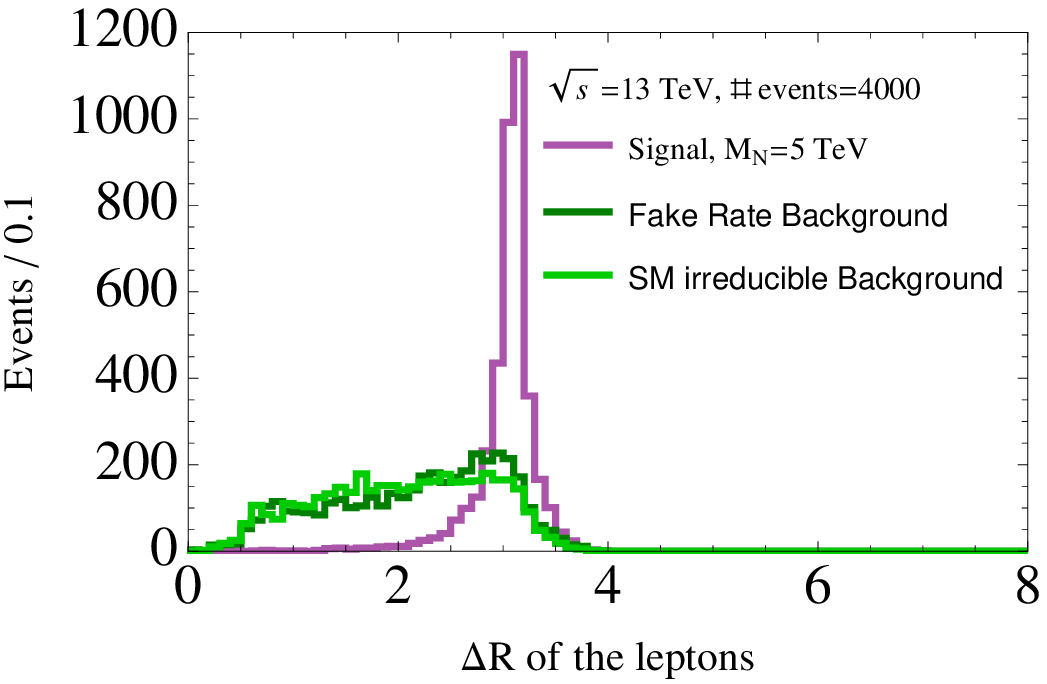}\subcaption{}
\end{subfigure}
\begin{subfigure}{3.0in}
\includegraphics[width=\textwidth]{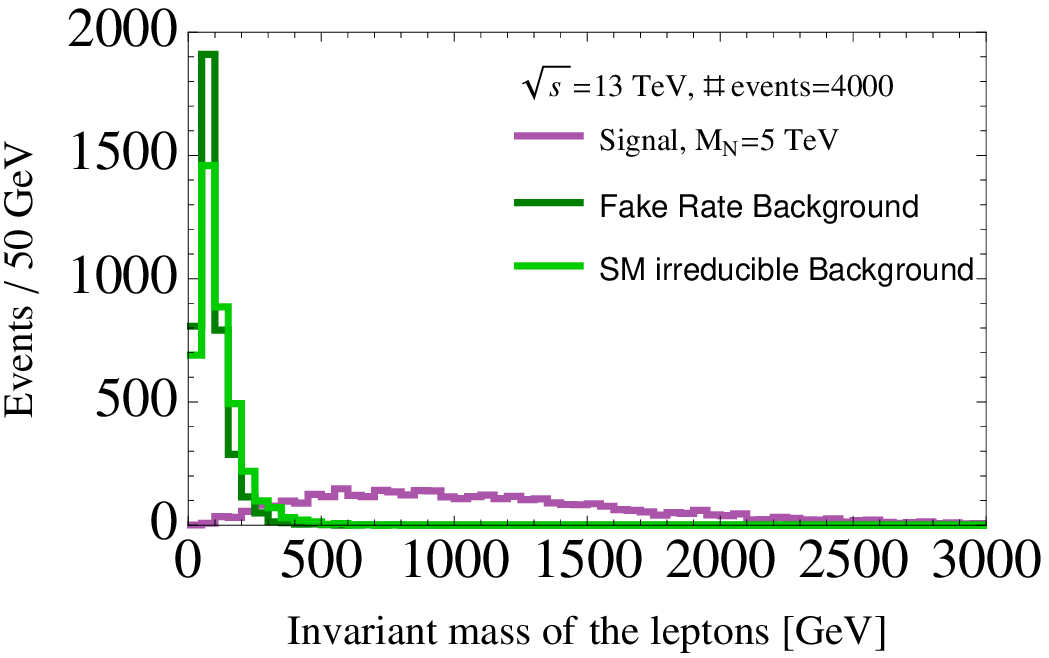}\subcaption{}
\end{subfigure}
\caption{\small $\Delta R_{ll}$ (left) and $m_{ll}$ (right) distributions for the signal and SM backgrounds using a heavy Majorana neutrino with mass $M_{N_{R}}=5$ TeV.}\label{fig:Fig13}
\end{figure}

The $T$-channel dominated signal is especially interesting since in the $M^{2}_{N_{R}}>|q|^{2}$ limit, the distribution of kinematical variables does not depend on the mass of the heavy Majorana neutrino. In addition, in this limit, the leptons are highly isotropic with $p_{T}$ values significantly higher than those of the SM background. This is shown in Figure~\ref{fig:Fig12}.
Finally, the lepton-pair separation, $\Delta R_{ll}$, and invariant mass distribution, $m_{ll}$, can also be used to reduce the background. The former is due to the fact that both leptons are back-to-back in the transverse plane. This is shown in Figure~\ref{fig:Fig13}.
Together with the default cuts introduced in Equation~(\ref{eq:Dcut}), we implement a series of additional cuts to extract a signal arising mainly from a $T$-channel dominated amplitude labeled T(s)13 and T(s)100 for $13$ and $100$ TeV machines respectively:
\begin{eqnarray}
\text{T13}&~~~~& m_{ll}> 500~\text{GeV},~p_{T,l}>200~\text{GeV},~|\eta_{l}| > 2, \nonumber \\
\text{T100}&~~~~& m_{ll}> 500~\text{GeV},~p_{T,l}>350~\text{GeV},~|\eta_{l}| > 2, \nonumber \\
\text{Ts13}&~~~~& m_{l_{1}l_{2},j_{1}j_{2}}>1500~\text{GeV},~~\epsilon_{j\to l}>10^{-5},\nonumber \\
\text{Ts100}&~~~~& m_{l_{1}l_{2},j_{1}j_{2}}>2500~\text{GeV},~~\epsilon_{j\to l}>10^{-5}.\label{eq:TchannelCuts}
\end{eqnarray}
While we emphasize the this strategy can yield large significances, it is very much dependent on the fake rate associated with the QCD background. That is, if the fake rate is higher, a tighter cut on the lepton $p_{T}$ must be used. Furthermore, the lepton separation, $\Delta R_{ll}$, is highly correlated with $m_{ll}$ and the lepton's transverse momentum; we observe that it only reduces the signal after all other cuts are applied.
\begin{table}[ht]\centering
 \tabcolsep 2.2 pt
\small
\begin{tabular}{|c|c|c|c|c|c|}
\hline
     &  & $e^{+}e^{+}\nu\nu~jj$ & $e^{+}\mu^{+}\nu\nu~jj$ & $\mu^{+}\mu^{+}\nu\nu~jj$ & $4j\cdot\epsilon^{2}_{j\to l}$  \\
\hline
\hline
Parton level             &      $\sigma$ (fb)       &	0.3927		&	0.7849	&	0.3927	& 1.79 \\
\hline
            &      $N$       &	1178		&	2354 	&	1178 	&  5370 \\
\hline
Detector level +~(\ref{eq:Dcut})            &        $\sigma$ (fb)      &		0.1187	&	0.2674	&	0.1187	&  0.471\\
\hline
           &         $N$    &	356		&	802	&	356	&  1410\\
\hline
\parbox[t]{0.40\textwidth}{ \centering T13 }               &      $\sigma$ (fb)        &		 $3.9\times10^{-6}$	 &	$7.8\times10^{-6}$	&	$<10^{-6}$	& $3.0\times10^{-5}$ \\
\hline
            &      $N$       &	$0.012$		&	$0.003$	&	$<0.003$	& $0.089$   \\
\hline
\parbox[t]{0.40\textwidth}{ \centering Ts13 }               &      $\sigma$ (fb)        &		 $<10^{-6}$	 &	$<10^{-6}$	&	$<10^{-6}$	& $2.69\times10^{-6}$ \\
\hline
            &      $N$       &	$<0.003$		&	$<0.003$	&	$<0.003$	& $0.0081$   \\
\hline
\end{tabular}
\caption{\small SM backgrounds at $13$ TeV. The event number is shown at a luminosity of $3000$ fb$^{-1}$. On the last two words we show the background reduction after applying the T13/Ts13 cut selection in Equation~(\ref{eq:TchannelCuts}).} \label{tab:Tab3}
\end{table}
\begin{table}[ht]\centering
 \tabcolsep 2.2 pt
\small
\begin{tabular}{|c|c|c|c|c|c|}
\hline
     &  & $e^{+}e^{+}\nu\nu~jj$ & $e^{+}\mu^{+}\nu\nu~jj$ & $\mu^{+}\mu^{+}\nu\nu~jj$ & $4j\cdot\epsilon^{2}_{j\to l}$  \\
\hline
\hline
Parton level             &      $\sigma$ (fb)       &	3.686		&	7.370	&	3.694	& 44.2\\
\hline
            &      $N$       &	11058		&	22110	&	11082	& 133000\\
\hline
Detector level +~(\ref{eq:Dcut})             &        $\sigma$ (fb)      &		0.962	&	2.148	&	1.189	& 11.6\\
\hline
           &         $N$    &	2887		&	6444 	&	3568 	& 35000\\
\hline
\parbox[t]{0.40\textwidth}{ \centering T100}               &      $\sigma$ (fb)        &	$<10^{-5}$	 	 &	$8.59\times10^{-4}$	&	$3.7\times10^{-6}$	& $7.4\times10^{-4}$ \\
\hline
            &      $N$       &	$<0.1$		&	$<0.44$	&	$0.11$	& $2.58$ \\
\hline
\parbox[t]{0.40\textwidth}{ \centering Ts100}               &      $\sigma$ (fb)        &	$<10^{-6}$	 	 &	$<2.0\times10^{-6}$	&	$<10^{-6}$	& $2.21\times10^{-4}$ \\
\hline
            &      $N$       &	$<0.003$		&	$<0.003$	&	$0.003$	& $0.663$ \\
\hline
\end{tabular}
\caption{\small  SM backgrounds at $100$ TeV. The event number is shown at a luminosity of $3000$ fb$^{-1}$. On the last two words we show the background reduction after applying the T100/Ts100 cut selection in Equation~(\ref{eq:TchannelCuts}).} \label{tab:Tab4}
\end{table}

\begin{figure}[ht]\centering
\begin{subfigure}{0.48\textwidth}
\includegraphics[width=\textwidth]{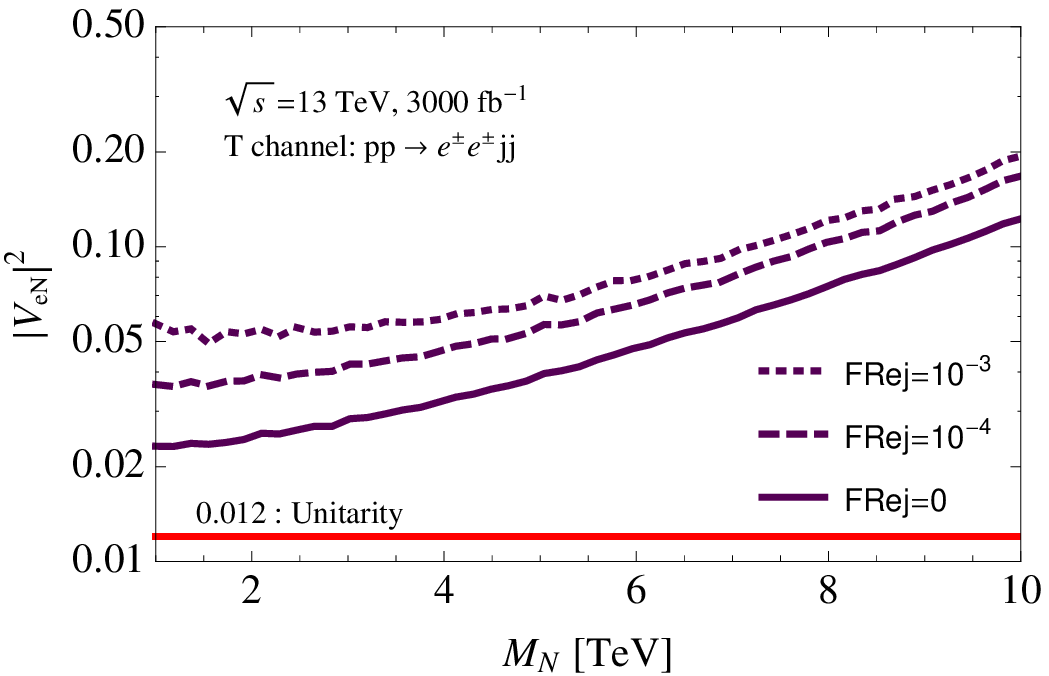}\subcaption{}
\end{subfigure}
\begin{subfigure}{0.48\textwidth}
\includegraphics[width=\textwidth]{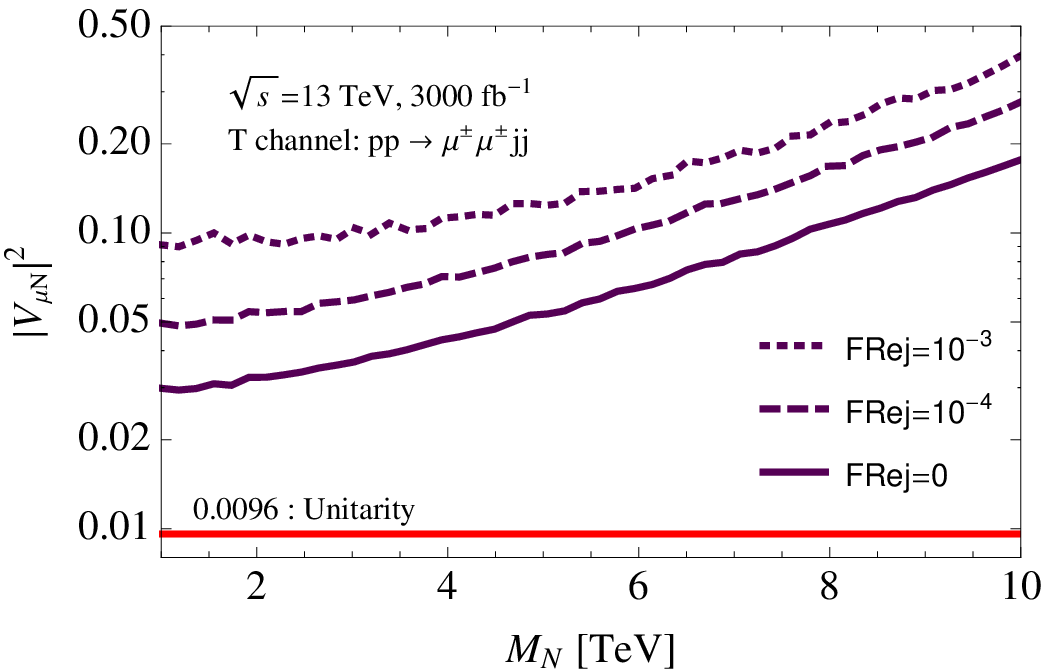}\subcaption{}
\end{subfigure}
\caption{\small The $95\%$ expected limit at $13$ TeV and $3000$ fb$^{-1}$ of integrated luminosity with varying fake rates on the QCD background in the same-sign electron channel (left) and same-sign muon channel (right). The red solid line indicates the bound from unitarity discussed in Section~\ref{subsec:constraints}. }\label{fig:Fig14}
\end{figure}
In Tables~\ref{tab:Tab3} and~\ref{tab:Tab4} we show the SM backgrounds after implementing the T(s)13 and T(s)100 cut selections. The Ts13 and Ts100 correspond to an additional cut on the $4$-particle final state invariant mass that reduces the QCD background by up to an order of magnitude in a $13$ TeV machine.
\begin{figure}[ht]\centering
\begin{subfigure}{0.48\textwidth}
\includegraphics[width=\textwidth]{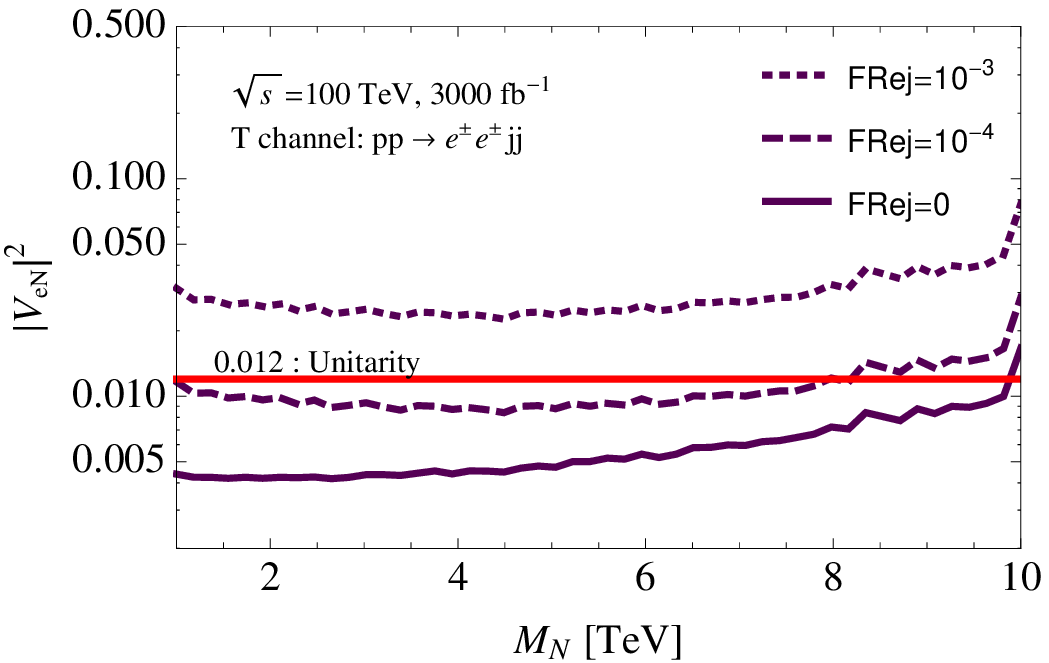}\subcaption{}
\end{subfigure}
\begin{subfigure}{0.48\textwidth}
\includegraphics[width=\textwidth]{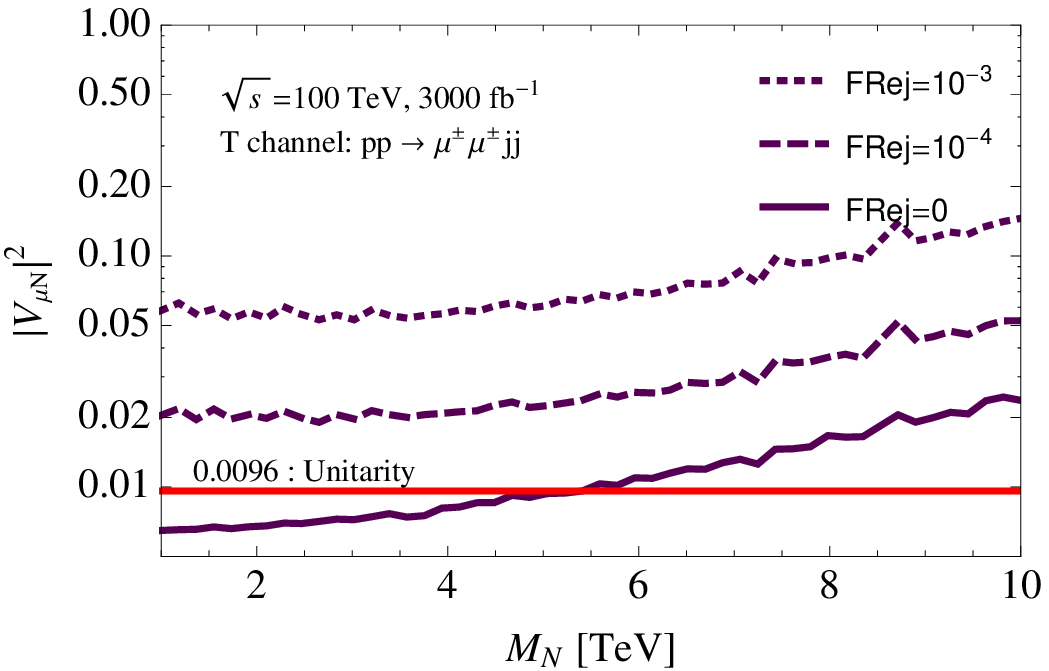}\subcaption{}
\end{subfigure}
\caption{\small The $95\%$ expected limit at $100$ TeV and $3000$ fb$^{-1}$ of integrated luminosity with varying fake rates on the QCD background in the same-sign electron channel (left) and same-sign muon channel (right). The red solid line indicates the bound from unitarity discussed in Section~\ref{subsec:constraints}.}\label{fig:Fig15}
\end{figure}
Using the cut flow discussed above, appropriate to extract a $T$-channel dominated signal; we observe that the SM irreducible background can be completely eliminated. In addition, implementing a fake rate of order ${\cal O}\left(10^{-5}\right)$ further suppresses the QCD background. However, the signal is proportional to the mixing angle raised to the fourth power, and the event yield is very small in the regions of small mixing angle. The yield can be enhanced with an additional cut on the $4$-particle final state invariant mass by one order of magnitude.  Using the analysis introduced in Section~\ref{subsec:Schannel} we find the $95\%$ confidence level fit using the $\chi$ variable defined in Equation~(\ref{eq:chiVar}) with varying mixing angle squared, $|V_{e4}|^{2},|V_{\mu 4}|^{2}$. We scan over the $p_{T}$ of the leptons, the pseudorapidity of the subleading jet, the same-sign lepton invariant mass and the $4$-particle invariant mass to extract the maximum value of $\chi$ leading to a $95\%$ expected limit on the mixing angle as a function of the Majorana neutrino masses. The scan is listed below:
\begin{itemize}
\item
$1.5<|\eta_{j_{2}}|<3$
\item
$0.1~\text{TeV}<p_{T}<0.7~\text{TeV}$
\item
$0.5~\text{TeV}<m_{l_{1}l_{2}}<1.~\text{TeV}$
\item
$0.7~\text{TeV}<m_{l_{1}l_{2}j_{1}j_{2}}<3.5~\text{TeV}$
\end{itemize}
 The results are shown in Figure~\ref{fig:Fig14} for a $13$ TeV collider and Figure~\ref{fig:Fig15} for $100$ TeV. In order to start probing couplings allowed by unitarity, in the large $M_{N_{R}}$ region, one will have to wait for a $100$ TeV collider and this highly depends on the ability to reduce the fake rate for both $\epsilon_{j\to e}$ and $\epsilon_{j\to \mu}$ below $\sim10^{-5}$. For very small fake rates, a $100$ TeV machine can probe Majorana masses up to $10$ TeV with couplings as low as $\epsilon_{e}\sim0.1$ in the same-sign electron channel.

\subsection{Left-Right Symmetric Model: Collider reach at $13$ and $100$ TeV}\label{subsec:LRchannel}
The presence of a charged gauged boson in the spectrum, associated with the SU(2)$_{\text{R}}$ gauge symmetry, leads to several phenomenological differences that makes a same-sign lepton analysis different from the basic $N_{R}$ extension of the SM. This is due to the fact that for gauge boson masses, $M_{W_{R}}$, above $1$ TeV, the $T$-channel contribution to the production cross section is strongly suppressed. This makes a $T$-channel dominated signal difficult to extract even with a $100$ TeV machine and $3000$ fb$^{-1}$ of integrated luminosity. Since we focus on the limit where  $M_{D}/M_{N_{R}}\ll1$, we can expect that $BR\left(N\to Wl\right)\to 0$, and are not able to use the SM charged gauge boson mass as a means to suppress the SM backgrounds. Thus, depending on the mass hierarchy between the Majorana neutrino and $W_{R}$ we can expect different processes to dominate the signal:
\begin{eqnarray}
M_{N_{R}}&<&M_{W_{R}}~~\sigma(pp\to l^{\pm}l^{\pm}~jj)_{\text{dominant}}=\sigma(pp\to W^{\pm}_{R})BR(W^{\pm}_{R}\to N_{R}~l^{\pm})BR(N_{R}\to l^{\pm}~jj) \nonumber \\
M_{N_{R}}&>&M_{W_{R}}~~\sigma(pp\to l^{\pm}l^{\pm}~jj)_{\text{dominant}}=\left\{
\begin{array}{lr}
(a)~\sigma(pp\to W^{\pm}_{R})BR(W^{\pm}_{R}\to l^{\pm}l^{\pm}~jj) \\
(b)~\sigma(pp\to W^{\pm *}_{R}\to N~l^{\pm})BR(N \to l^{\pm}~jj) \\
\end{array} \right. \label{eq:3.11} \nonumber \\
\end{eqnarray}

\begin{figure}[ht]\centering
\begin{subfigure}{0.48\textwidth}
\includegraphics[width=\textwidth]{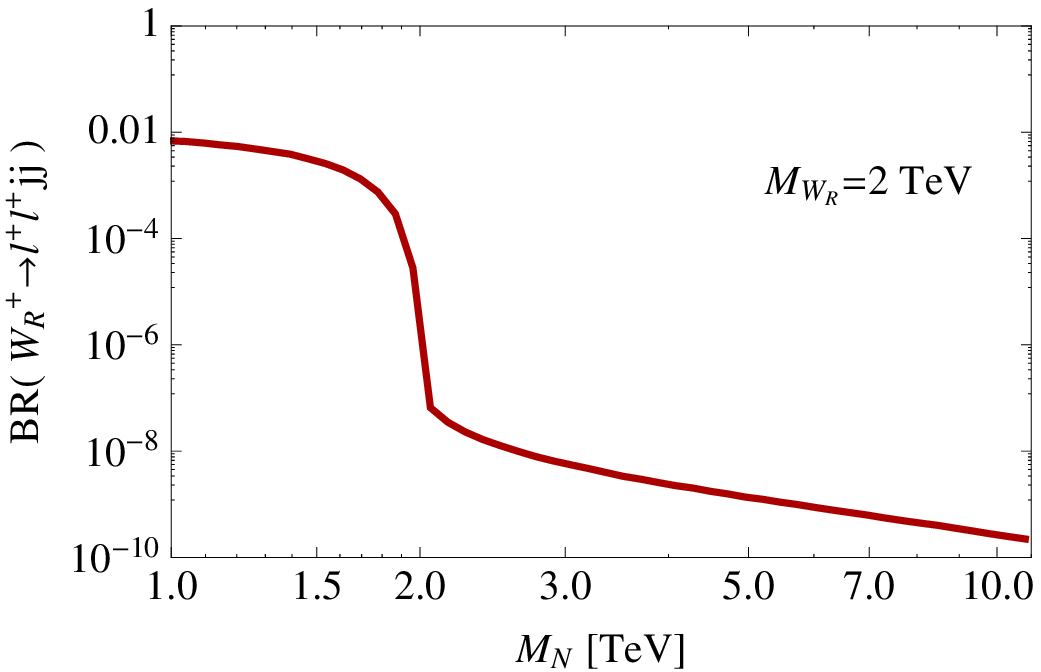}\subcaption{}
\end{subfigure}
\begin{subfigure}{0.48\textwidth}
\includegraphics[width=\textwidth]{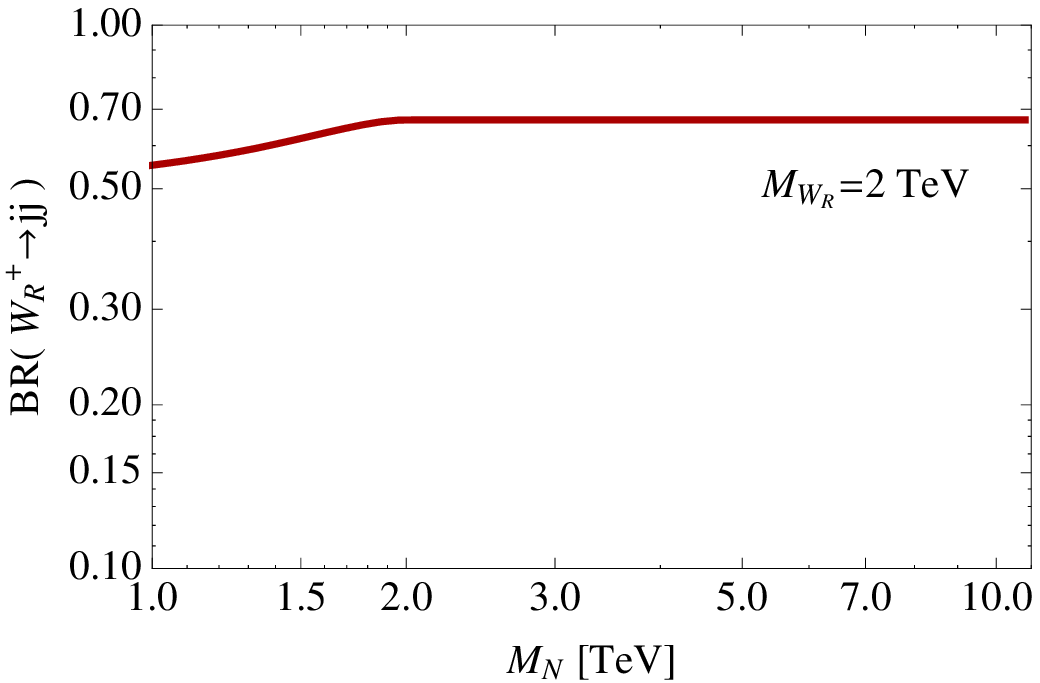}\subcaption{}
\end{subfigure}
\caption{\small Branching ratio of $W^{\pm}_{R}\to l^{\pm}l^{\pm}~jj$ (left) and $W^{\pm}\to jj$ (right) with $M_{W_{R}}=2$ TeV. }\label{fig:Fig16}
\end{figure}
The search strategy will strongly depend on whether $W_{R}$ can be produced on-shell. In particular, a same-sign lepton plus two jets final state will be suppressed by a $4$-body phase space factor in the region where $M_{W_{R}}<M_{N_{R}}$. We can see this in Figure~\ref{fig:Fig16}(a) for $M_{W_{R}}=2$ TeV. However, from Figure~\ref{fig:Fig16}(b), we can see that the branching ratio of $W_{R}$ into two jets is mostly unaffected. Therefore, for $M_{N_{R}}<M_{W_{R}}$, the $p_{T}$ and $\eta$ of the final state particles will highly depend on the $p_{T}$ and $\eta$ of the $W_{R}$ on its rest frame. However, for $M_{N_{R}}>M_{W_{R}}$, resonances can be reconstructed depending on the phase space:
\begin{itemize}
\item
$(a)$~$M_{W_{R}}=M_{l_{1}l_{2},j_{1}j_{2}}$,
\item
$(b)$~$M_{W_{R}}=M_{j_{1}j_{2}},~M_{N_{R}}=M_{l_{1},j_{1}j_{2}}$.
\end{itemize}
\begin{figure}[ht]\centering
\begin{subfigure}{0.48\textwidth}
\includegraphics[width=\textwidth]{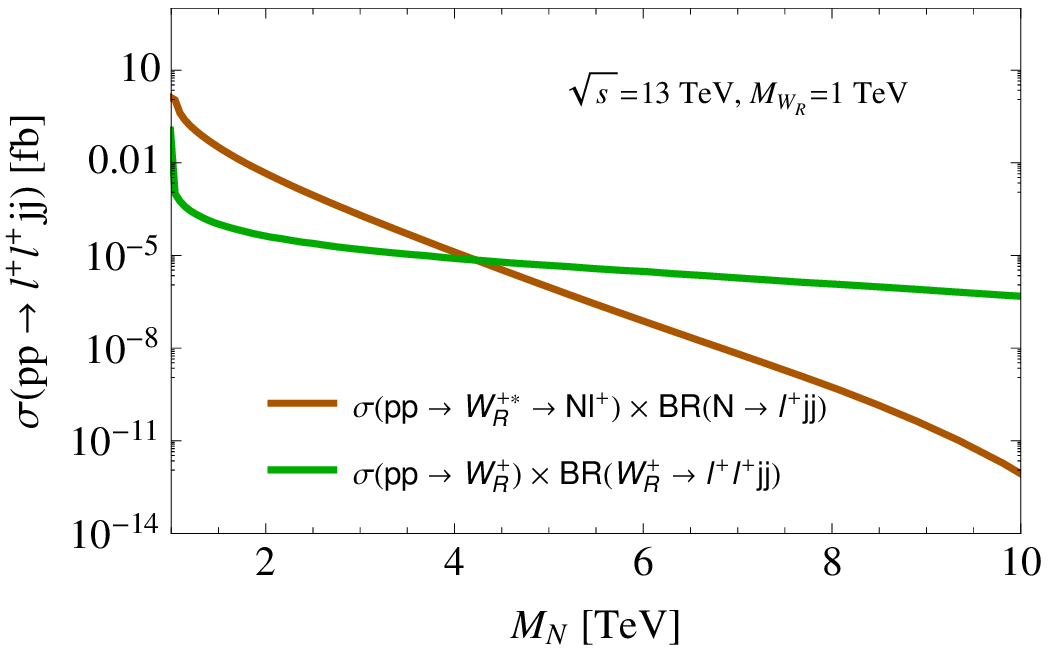}\subcaption{}
\end{subfigure}
\begin{subfigure}{0.48\textwidth}
\includegraphics[width=\textwidth]{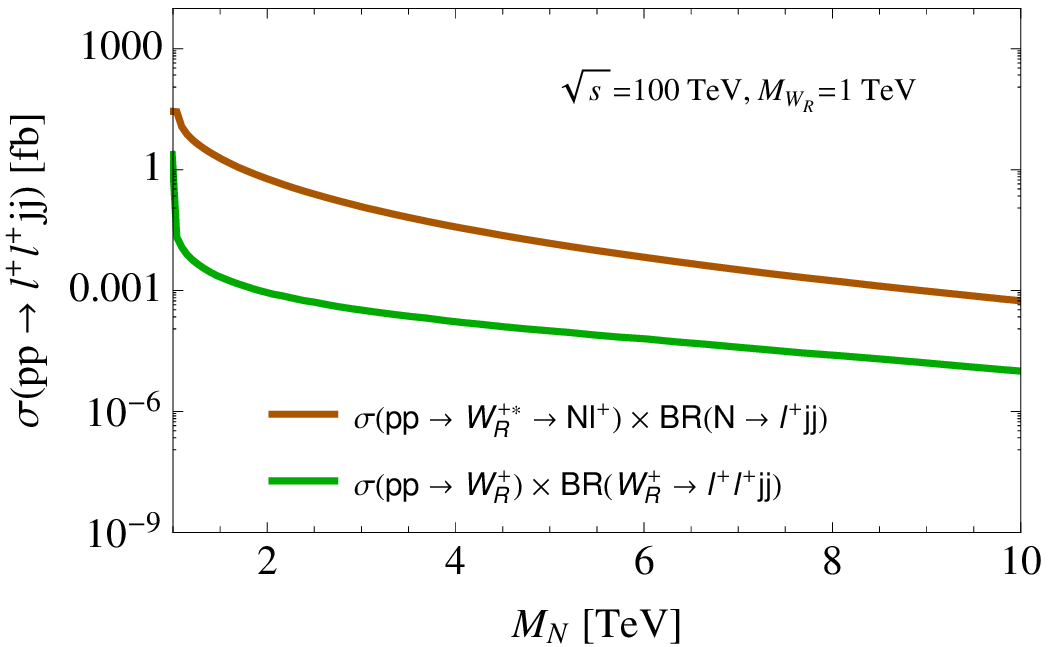}\subcaption{}
\end{subfigure}
\begin{subfigure}{0.48\textwidth}
\includegraphics[width=\textwidth]{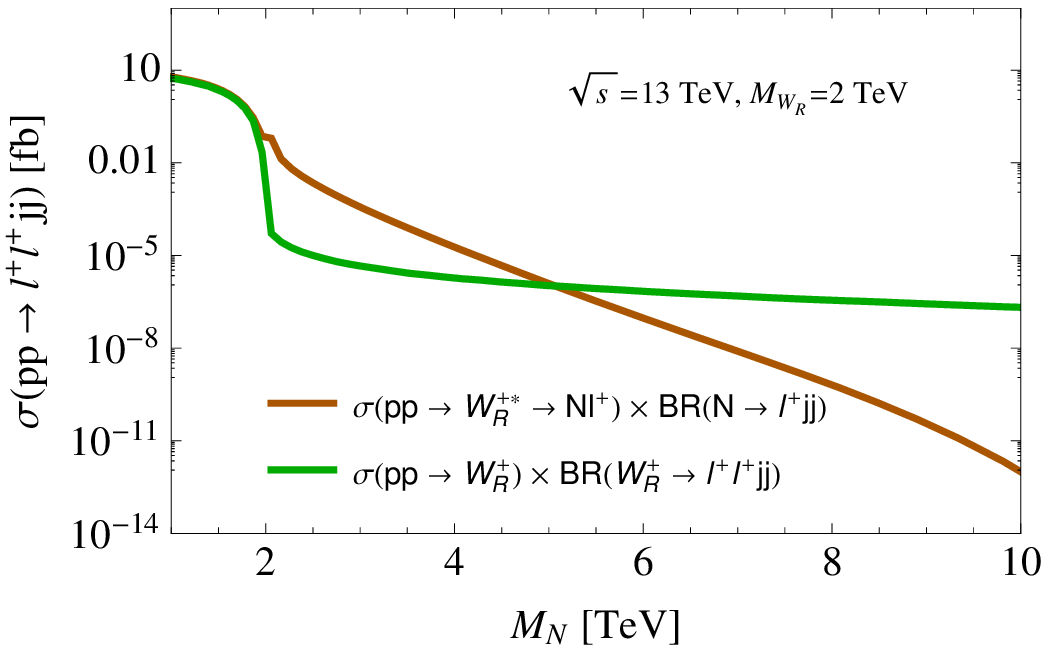}\subcaption{}
\end{subfigure}
\begin{subfigure}{0.48\textwidth}
\includegraphics[width=\textwidth]{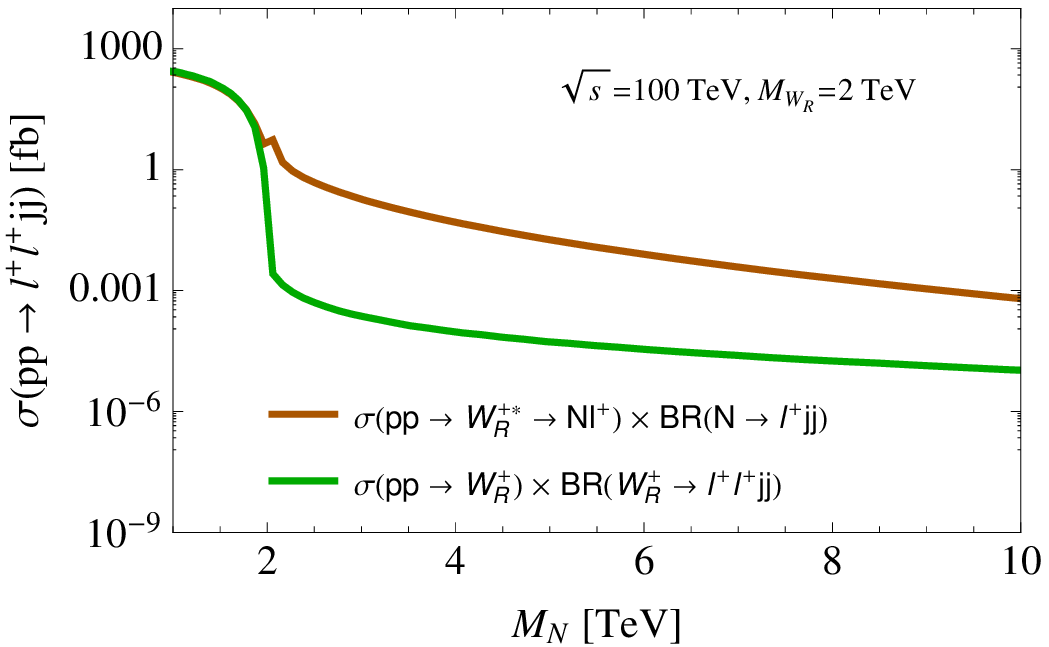}\subcaption{}
\end{subfigure}
\begin{subfigure}{0.48\textwidth}
\includegraphics[width=\textwidth]{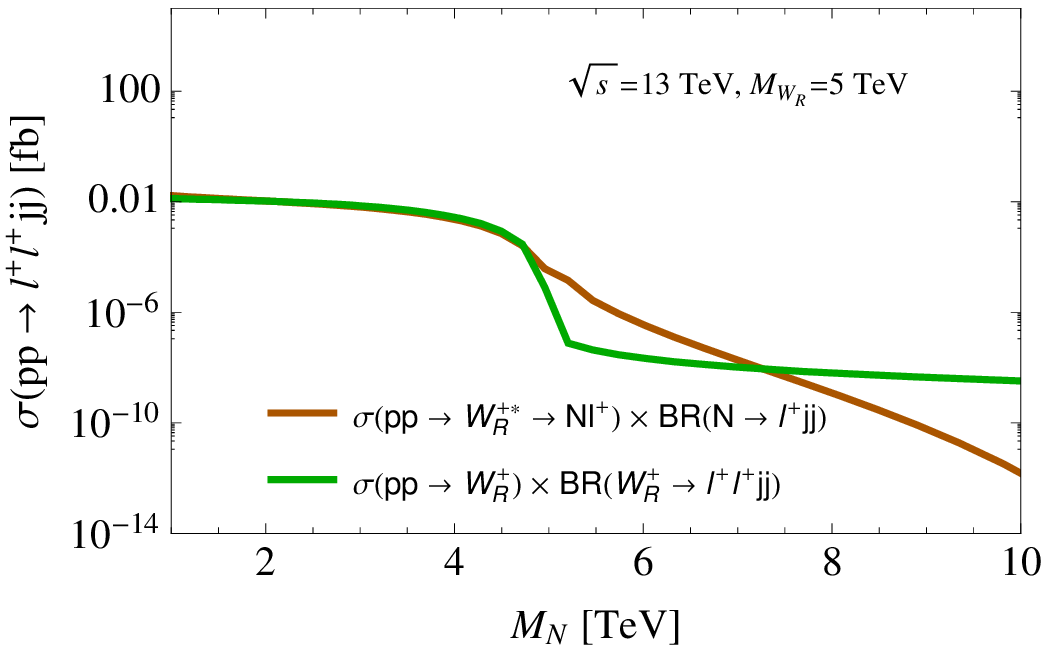}\subcaption{}
\end{subfigure}
\begin{subfigure}{0.48\textwidth}
\includegraphics[width=\textwidth]{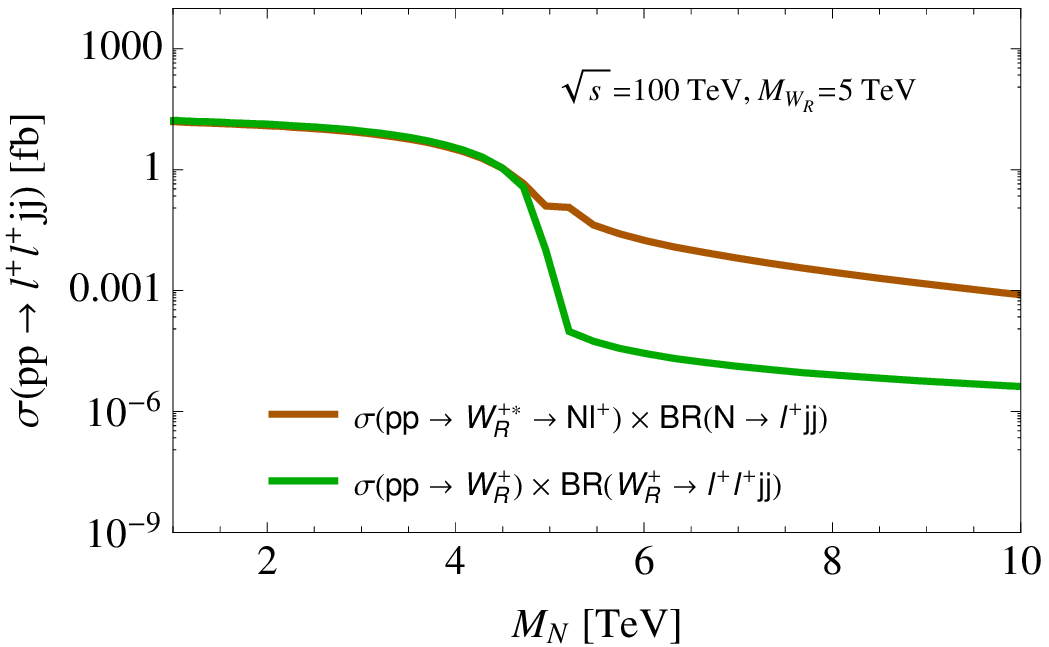}\subcaption{}
\end{subfigure}
\caption{\small $pp\to l^{+}l^{+}~jj$ cross section for varying $M_{N_{R}}$ at $13$ TeV (left) and $100$ TeV (right) center of mass energies. The top, middle and lower panels correspond to masses $M_{W_{R}}=1,2,5$ TeV respectively.}\label{fig:Fig17}
\end{figure}
To see which resonances can be used to better enhance a same-sign lepton signal, we analyze how (a) and (b) in Equation~(\ref{eq:3.11}) behave as a function of $M_{N_{R}}$ and where the on-shell production of the right-handed gauge boson dominates. This is shown in Figure~\ref{fig:Fig17} where on the left column we plot the production cross section at $13$ TeV as a function of $M_{N_{R}}$ for $M_{W_{R}}$ values of $1,2$ and $5$ TeV respectively while results for $100$ TeV are shown on the right column. In the former we observe a small region for $M_{N_{R}}>M_{W_{R}}$ where we cannot use the $lj_{1}j_{2}$ invariant mass to reconstruct the Majorana mass while in the latter, the (b) production mode always dominates. Thus, for a $100$ TeV machine the invariant mass of the $lj_{1}j_{2}$ system can always be used to reconstruct the Majorana neutrino mass for both $M_{N_{R}}<M_{W_{R}}$ and $M_{N_{R}}>M_{W_{R}}$.

After a fast detector simulation, most of the leptons are misidentified as jets. With Delphes, at $13$ TeV, we find a signal survival rate of about $10-25\%$, while for a $100$ TeV machine the survival rate is below $10\%$ for Majorana masses above $\sim 3$ TeV. We note that this rate is unreasonably small and a more complete analysis on the simulation of different detector types should be used. In our study we conclude that is difficult to probe our simplified version of the left-right symmetric model with $M_{W_{R}}>3,M_{N_{R}}>1$ TeV at a $13$ TeV machine and $M_{W_{R}},M_{N_{R}}>8$ TeV at $100$ TeV unless we significantly reduce the SM background. In what follows we only study a same-sign electron signal. A proper study of the muon chamber ID performance can make a same-sign muon signal just as significant and the analysis would proceed in the same way.

\begin{figure}[ht]\centering
\begin{subfigure}{0.48\textwidth}
\includegraphics[width=\textwidth]{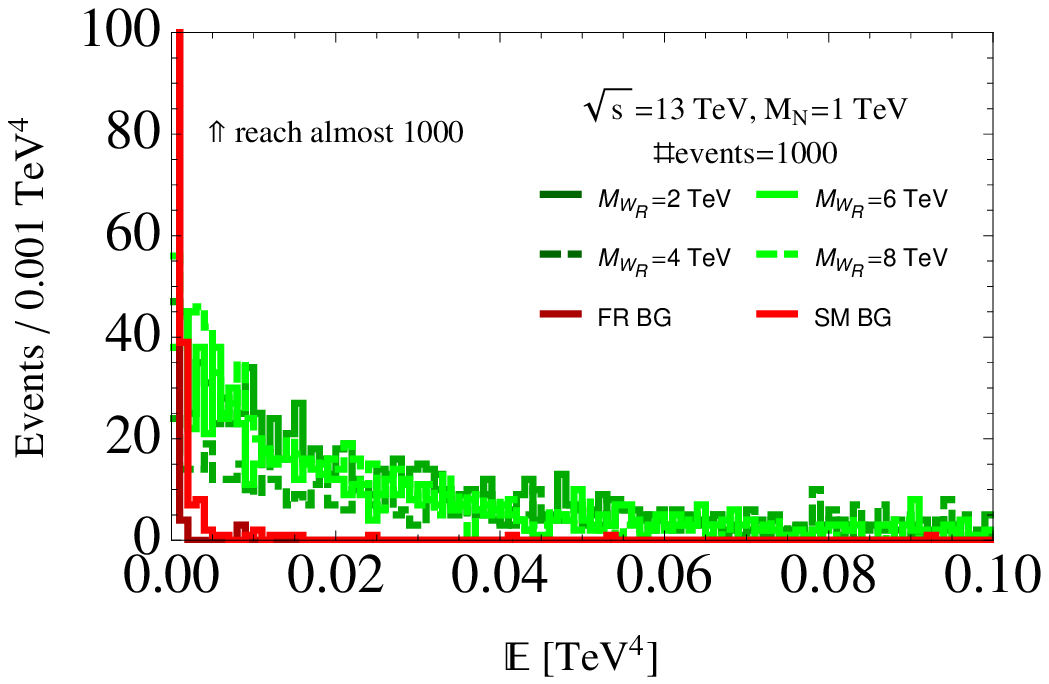}\subcaption{}
\end{subfigure}
\begin{subfigure}{0.48\textwidth}
\includegraphics[width=\textwidth]{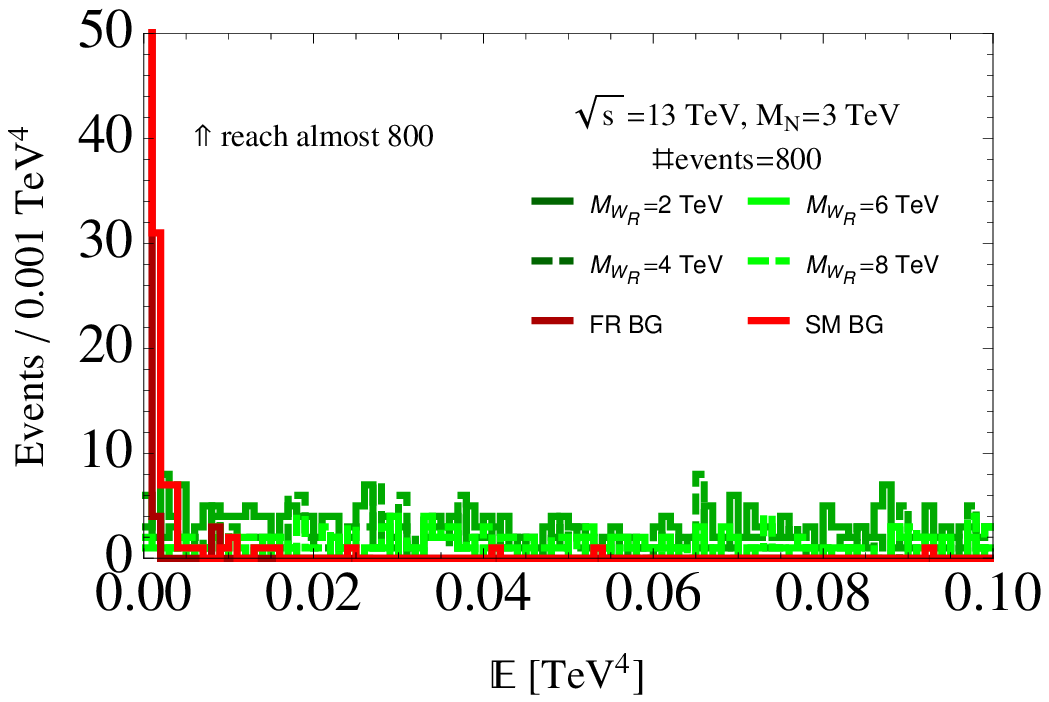}\subcaption{}
\end{subfigure}
\begin{subfigure}{0.48\textwidth}
\includegraphics[width=\textwidth]{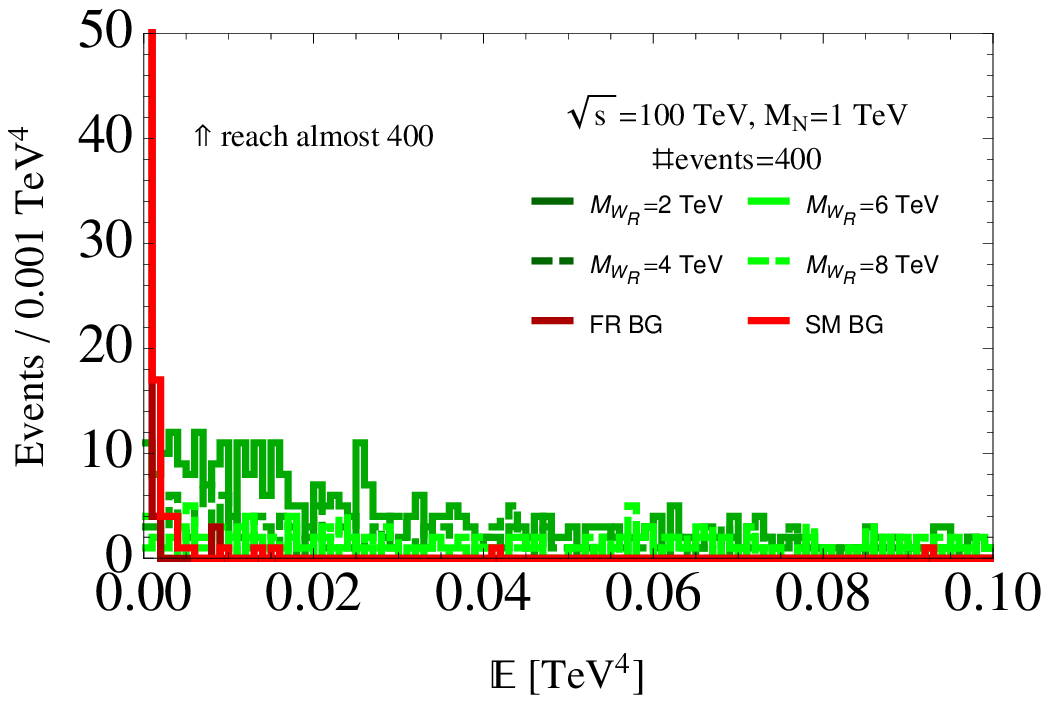}\subcaption{}
\end{subfigure}
\begin{subfigure}{0.48\textwidth}
\includegraphics[width=\textwidth]{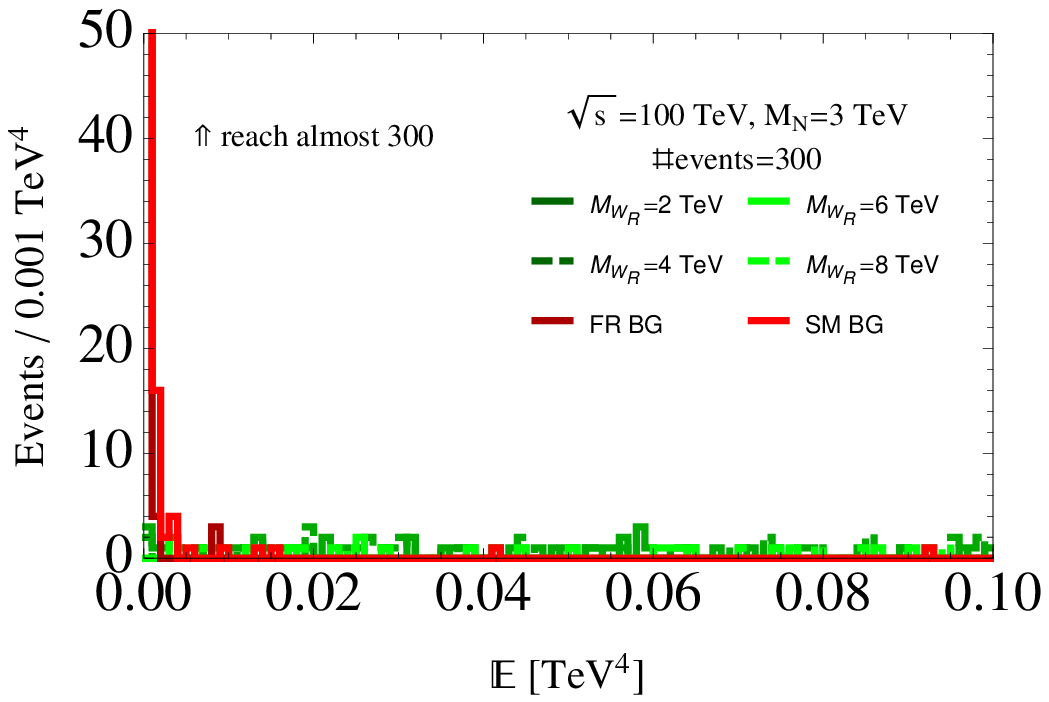}\subcaption{}
\end{subfigure}
\caption{\small The $\mathbb{E}$ distribution of all final state particles with $M_{N_{R}}=1$ TeV (left) and $M_{N_{R}}=3$ TeV. The top panel corresponds to center of mass energies of $13$ TeV while the bottom panel to $100$ TeV. }\label{fig:Fig18}
\end{figure}
In addition to the $3$- and $4$-particle invariant masses discussed above, we implement an additional $4$-particle Lorentz invariant kinematical variable that we find extremely useful to suppress the SM background, details which can be found in Appendix~\ref{app:AppB}. We label this variable by $\mathbb{E}$ and define it as
\begin{equation}
\mathbb{E}=\epsilon_{\mu\nu\sigma\rho}P^{\mu}_{1}P^{\nu}_{2}P^{\sigma}_{3}P^{\rho}_{4},
\label{eq:ELorentz}\end{equation}
\begin{table}[ht]\centering
 \tabcolsep 2.2 pt
\small
\begin{tabular}{|c||c|c||c|c|}
\hline
$\mathbb{E}$ [GeV$^{4}$] & $e^{+}e^{+}\nu\nu~jj$ ($13$ TeV) & Fake ($13$ TeV) & $e^{+}e^{+}\nu\nu~jj$ Fake ($100$ TeV) &  Fake ($100$ TeV)\\
\hline
\hline
$10^{9}$     		& $3.367\%$	&	$0.769\%$	&	$10.954\%$	& $3.327\%$ \\
\hline
$5\times10^{9}$       & $0.983\%$	& 	$0.100\%$	&	$2.358\%$	& $0.746\%$\\
\hline
$10^{10}$      		& $0.350\%$	&	$0.031\%$	&	$1.058\%$	& $0.350\%$\\
\hline
$5\times10^{10}$     & $0.023\%$	&	 $0.0027\%$	&	$0.119\%$ 	& $0.062\%$\\
\hline
$10^{11}$        		& $<0.003\%$	&	$<0.0002\%$	&	$0.027\%$	& $0.023\%$\\
\hline
\end{tabular}
\caption{\small The cut efficiency for the $\mathbb{E}$ variable on the leading SM background for $13 $ and $100$ TeV center of mass energies.  } \label{tab:Tab5}
\end{table}
where $P_{1,2,3,4}$ denote the four-momentum of the different particles. The expression can be simplified if we define the $z$-axis to be the direction of particle $\vec{p}_{1}$. The effectiveness of this variable to reduce the SM background is depicted in Figure~\ref{fig:Fig18} and a qualitatively analysis is shown in Table~\ref{tab:Tab5}.

In order to enhance a left-right symmetric signal over the SM background we implement a series of cuts on the $p_{T}$ of the leading lepton, the $4$-particle final state invariant mass, $m_{l_{1}l_{2},j_{1}j_{2}}$ and $\mathbb{E}$. The cuts are chosen to maximize the statistical estimator, $S/\sqrt{S+B}$. Furthermore, we use mass relations between $M_{N_{R}}$ and $M_{W_{R}}$ and two different fake rates ($10^{-4},10^{-5}$). In the region where $M_{N_{R}}>M_{W_{R}}$ , the width of $N_{R}$ to $W_{R}$ and a lepton is given by
\begin{equation}
\Gamma(N\to lW_{R})\approx \frac{g^{2}_{R}}{8\pi}|V_{lN}|^{2}M_{N_{R}}\left[\frac{M^{2}_{N_{R}}}{M^{2}_{W_{R}}}+1\right],
\end{equation}
and for $M_{N_{R}}<M_{W_{R}}$ they are given by
\begin{eqnarray}
\Gamma(W_{R}\to j_{1}j_{1})&\approx&\frac{g^{2}_{R}}{8\pi}|V_{j_{1}j_{2}}|^{2}\frac{3}{2}M_{W_{R}}, \nonumber \\
\Gamma(W_{R}\to Nl)&\approx&\frac{g^{2}_{R}}{8\pi}|V_{lN}|^{2}\frac{M^{2}_{W_{R}}-M^{2}_{N_{R}}}{2M_{W_{R}}}\left[3-2\frac{M^{2}_{N_{R}}}{M^{2}_{W_{R}}}-\frac{M^{4}_{N_{R}}}{M^{4}_{W_{R}}}\right].
\end{eqnarray}
Using figure~\ref{fig:Fig17} we see that at $13$ TeV, we may focus on the region where $M_{W_{R}}>M_{N_{R}}$ and on-shell production of the SU(2)$_{\text{R}}$ gauge boson, while with a $100$ TeV collider we may probe both mass hierarchies with on-shell $W_{R}$ or $N_{R}$ production.

\begin{figure}[ht]\centering
\begin{subfigure}{0.48\textwidth}
\includegraphics[width=\textwidth]{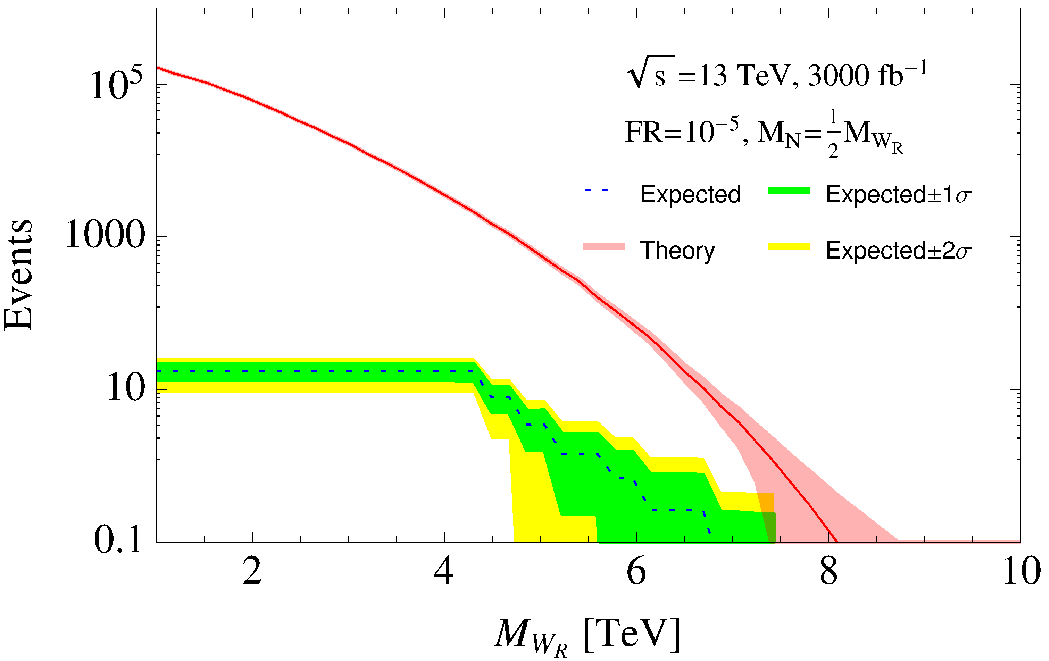}\subcaption{}
\end{subfigure}
\begin{subfigure}{0.48\textwidth}
\includegraphics[width=\textwidth]{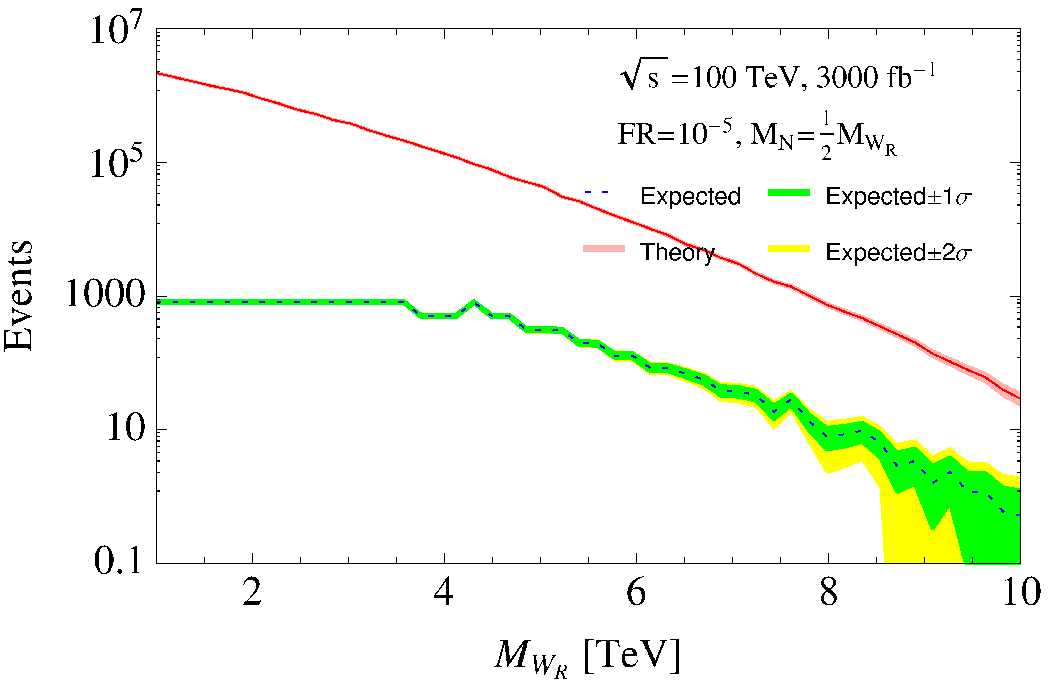}\subcaption{}
\end{subfigure}
\begin{subfigure}{0.48\textwidth}
\includegraphics[width=\textwidth]{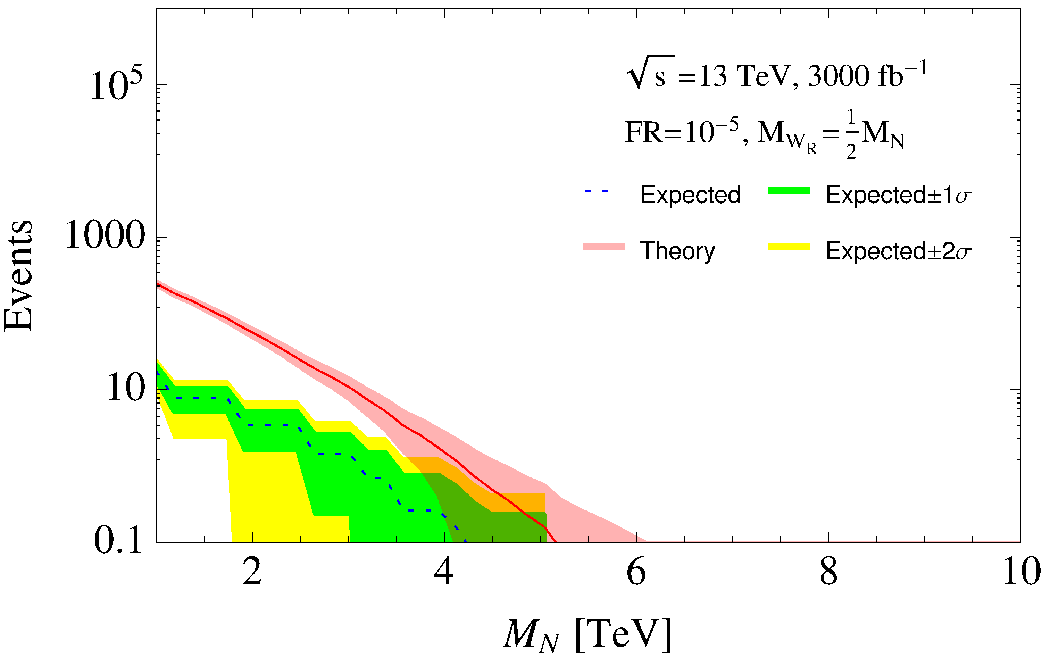}\subcaption{}
\end{subfigure}
\begin{subfigure}{0.48\textwidth}
\includegraphics[width=\textwidth]{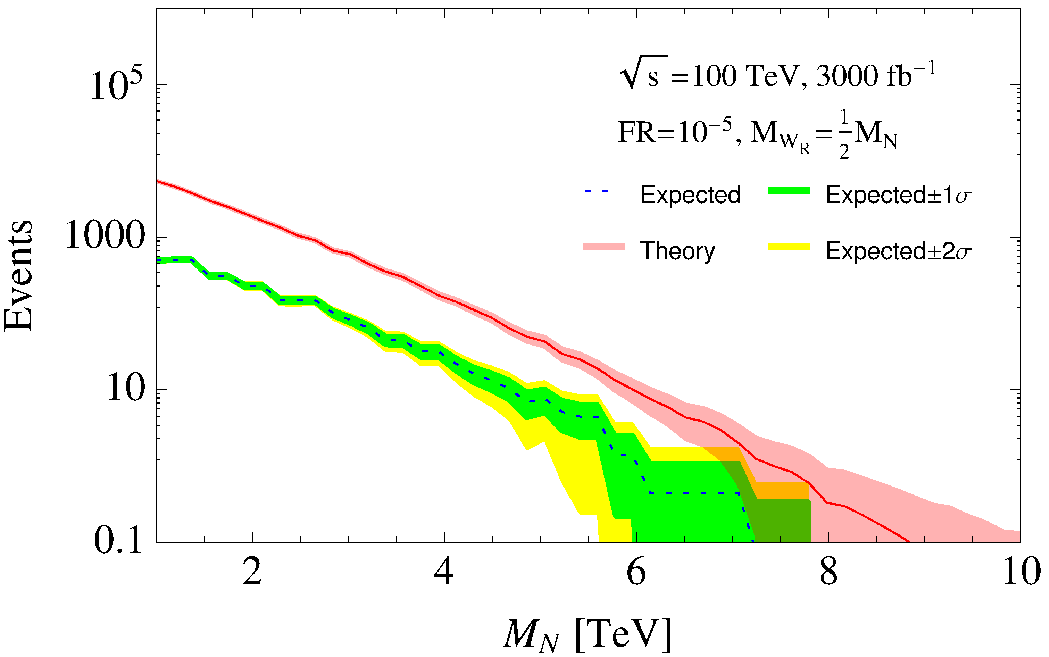}\subcaption{}
\end{subfigure}
\caption{\small Expected background and theory prediction with $M_{N_{R}}=\frac{1}{2}M_{W_{R}}$ (top panel) at $13$ TeV on the left and $100$ TeV on the right. The lower panel corresponds to $M_{W_{R}}=\frac{1}{2}M_{N_{R}}$.   }\label{fig:Fig19}
\end{figure}
\begin{figure}[ht]\centering
\begin{subfigure}{0.48\textwidth}
\includegraphics[width=\textwidth]{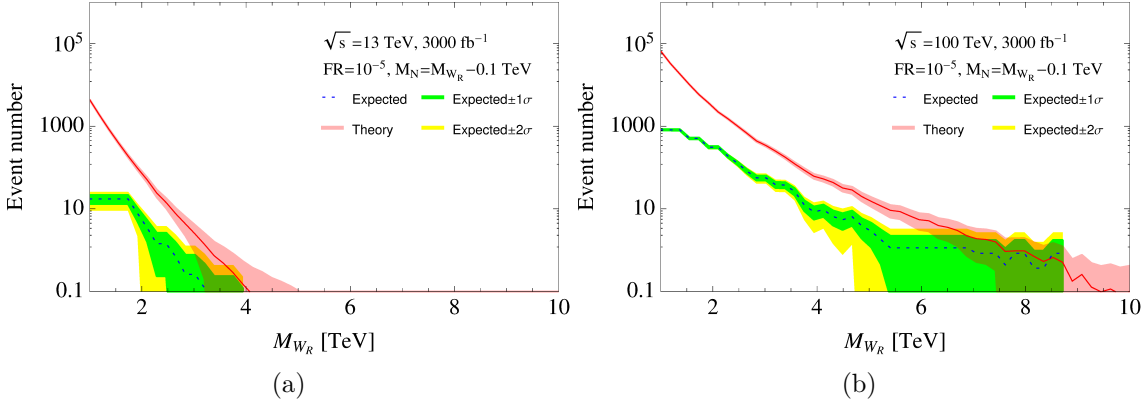}\subcaption{}
\end{subfigure}
\begin{subfigure}{0.48\textwidth}
\includegraphics[width=\textwidth]{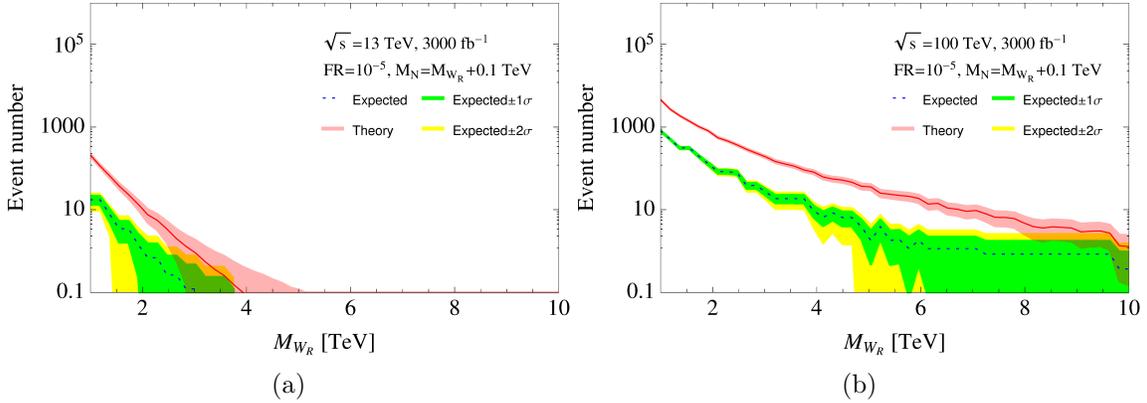}\subcaption{}
\end{subfigure}
\caption{\small  Expected background and theory prediction with $M_{N_{R}}=M_{W_{R}}-0.1~\text{TeV}$ (top panel) at $13$ TeV on the left and $100$ TeV on the right.}\label{fig:Fig20}
\end{figure}
\begin{figure}[ht]\centering
\begin{subfigure}{0.48\textwidth}
\includegraphics[width=\textwidth]{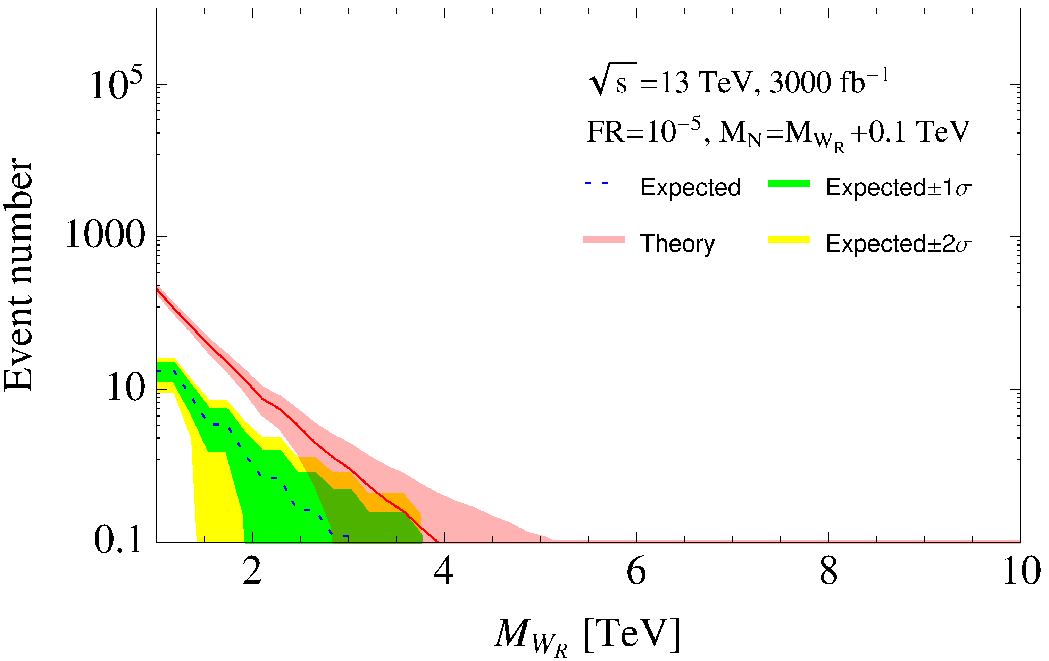}\subcaption{}
\end{subfigure}
\begin{subfigure}{0.48\textwidth}
\includegraphics[width=\textwidth]{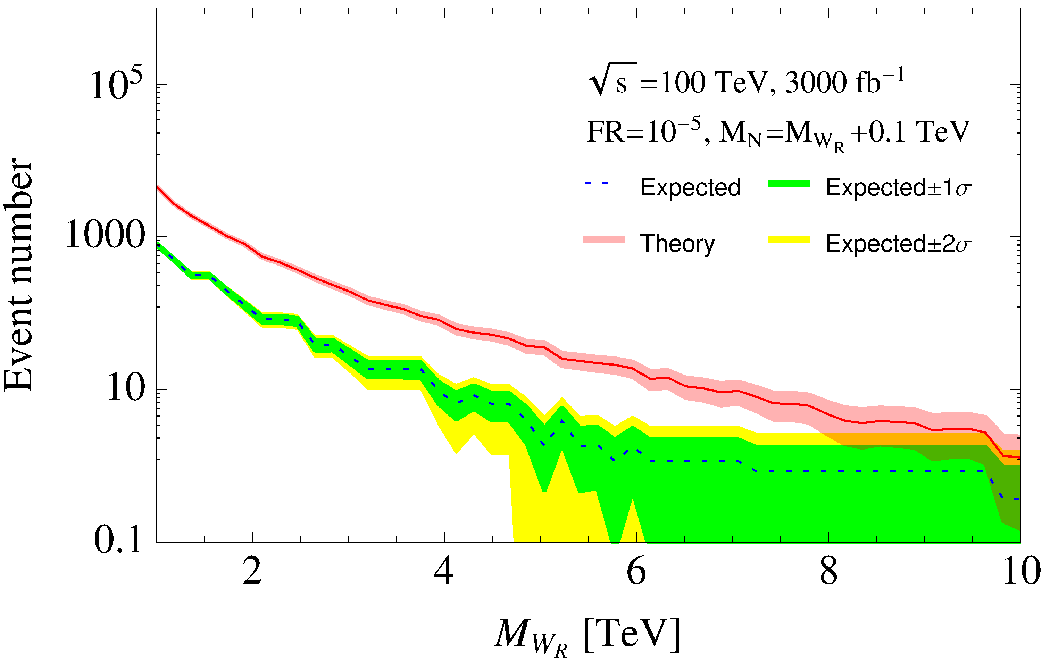}\subcaption{}
\end{subfigure}
\caption{\small Expected background and theory prediction with $M_{N_{R}}=M_{W_{R}}+0.1~\text{TeV}$ (top panel) at $13$ TeV on the left and $100$ TeV on the right.  }\label{fig:Fig21}
\end{figure}

We calculate the collider reach in the non-degenerate region for $M_{W_{R}}>M_{N_{R}}$ with $M_{N_{R}}=\frac{1}{2}M_{W_{R}}$ (top panel) and for $M_{W_{R}}<M_{N_{R}}$ with $M_{W_{R}}=\frac{1}{2}M_{N_{R}}$ (bottom panel). We emphasize that our choice of masses is not theoretically motivated and we use it to illustrate the significance of a same-sign lepton signal. The total number of events before cuts depends on $BR(W_{R}\to Nl)\times BR(N_{R}\to ljj)$ for $M_{W_{R}}>M_{N_{R}}$ and $BR(N_{R}\to ljj)$ for $M_{N_{R}}>M_{W_{R}}$; the later which is close to $100\%$. These allows us to scale the cross sections accordingly except for when the two masses are very degenerate, since in this case the leptons in the final state will be very compressed. We present only the same-sign electron channel and include only the statistical error using the least conservative fake rate of $\epsilon_{j\to l}=10^{-5}$. The same-sign muon channel is suppressed with respect to the electron channel since within Delphes, the muon reconstruction efficiency is lower. A realistic muon-chamber simulation may lead to better results for the same-sign muon final state. Results were obtained from a scan on the $p_{T}$ of the leptons, the $4$-particle final state invariant mass and the $\mathbb{E}$ variable used to obtain the highest significance possible with $3000$ fb$^{-1}$ at center of mass energies of $13$ and $100$ TeV. Furthermore, values for $V_{eN_{R}},V_{\mu N_{R}}=1$ are used. The scan is listed below:
\begin{itemize}
\item
$10^{-4}~\text{TeV}^{4}<\mathbb{E}<10^{-1}~\text{TeV}^{4}$
\item
$0.1~\text{TeV}<p_{T}<1.~\text{TeV}$
\item
$0.7~\text{TeV}<m_{l_{1}l_{2}j_{1}j_{2}}<4.5~\text{TeV}$
\end{itemize}
It is interesting to see that even a $13$ TeV collider can probe Majorana neutrino masses up to $6$ TeV and gauge boson masses up to $8$ TeV. In addition, the degenerate region is also very interesting and we calculate the reach for $M_{N_{R}}=M_{W_{R}}-0.1$ TeV and $M_{N_{R}}=M_{W_{R}}+0.1$ TeV. These are shown in Figures~\ref{fig:Fig20} and~\ref{fig:Fig21} respectively. In all four cases, larger values of $\mathbb{E}$ are required to suppress the SM background for large $W_{R}$ and $N_{R}$ masses. In addition, we observe that for a $13$ TeV collider, a flat $p_{T}$ cut at around $0.1$ TeV is sufficient; while the $4$-particle invariant mass cut increases with both the mass and the collider energy. We recommend that the experimental collaborations explore cuts on the $p_{T}$ of the leading lepton above $0.1$ TeV and on the total invariant mass between $1$ and $2.5$ TeV at the LHC's run II, and to also consider implementing the $\mathbb{E}$ variable. However, we point out that we have assumed that the PDFs implemented in MadGraph are adequate for a $100$ TeV study. An in-depth study is necessary, especially since the plans for the construction of a $100$ TeV machine appears to be gaining enough momentum.
\section{Discussion}\label{sec:discussion}

In this study we have carried out an in depth analysis of two simplified scenarios that give rise to a heavy right-handed Majorana neutrino. In particular we have focused on a same-sign lepton final state that can be probed at the LHC Run II and a future $100$ TeV hadron collider. Our work compliments existing analyses where the major focus has been on on-shell production of the Majorana neutrino with masses below $1$ TeV. We have extended our analysis to masses above $1$ TeV by determining a set of optimal cuts that can be used to extract a $T$-channel dominated same-sign lepton signal. In particular, we observe that the same-sign lepton invariant mass is an excellent variable to suppress the SM irreducible background while the $4$-particle final state invariant mass can be used to significantly suppress the QCD background using a fake rate of order ${\cal O}\left(10^{-5}\right)$. With the set of cuts given in Equation~(\ref{eq:TchannelCuts}) we are able to probe the same-sign electron channel beyond bounds from lepton unitarity with $3000$ fb$^{-1}$ of data at $13$ TeV. Furthermore, we observe that a significant same-sign muon signal can be probe, beyond limits from lepton unitarity, with a $100$ TeV machine.

In addition, we have analyzed the sensitivity to a simplified version of the left-right symmetric model, where in addition to a heavy Majorana neutrino, right-handed charged currents mediate lepton violating interactions. Given this additional degree of freedom, we conduct a study based on whether $W_{R}$ or $N_{R}$ can be produced on-shell, and we optimize the collider study to extract a signal in different regions of phase space. Furthermore; our analysis goes beyond and compliments the experimental searches by CMS and ATLAS~\cite{Khachatryan:2014dka,ATLAS:2012ak} in the region where $M_{N_{R}}>1$ TeV. We find an interesting $4$-particle Lorentz invariant quantity, $\mathbb{E}$, introduced in Equation~(\ref{eq:ELorentz}), that can be used to suppress the background in all regions of phase space. This variable peaks in the presence of heavy beyond-the-SM particles. The analysis is carried out for a mass relation between $N_{R}$ and $W_{R}$ and find that in regions where $M_{N_{R}}=M_{W_{R}}/2$ the LHC 13 TeV with $3000$ fb$^{-1}$ can reach heavy gauge boson masses up to $\sim8$ TeV, unlike the degenerate region, where the reach is only up to $4$ TeV. The inverse hierarchy where $M_{N_{R}}>M_{W_{R}}$ is more difficult to probe since the on-shell production is phase space suppressed, however, the reach is up to $M_{N_{R}}\sim5$ TeV in the region where $M_{W_{R}}=M_{N_{R}}/2$. In addition, we have not commented on the reach of the $e\mu~jj$ channel, which unlike the single Majorana extension of the SM, is much more promising in the left-right symmetric model, due to the additional suppression from the ratio $\left(M_{W}/M_{W_{R}}\right)^{4}$ in the $\mu\to e\gamma$ branching ratio given in Equation~(\ref{eq:muegLR}). We understand that the analysis in~\cite{Abada:2014cca} shows that FCC-ee(TLEP)~\cite{FCC-ee} would have a better constrain on the flavor violating processes through rare Z decay measurement. Being a leptonic collider it has the advantage of lower backgrounds but it is sensitive to lower mass heavy states.  On the other hand  hadronic colliders can probe higher masses as we have shown, but require the new couplings to be relatively large. Hence, this study complements the FCC-ee study of heavy neutrinos. Optimistically, if a signal is found for a heavy Majorana neutrino in the mass range we have examined it will mean that simple Type I seesaw with very small mixings is disfavored. This will point to either more structure in the heavy neutrino sector and/or a rich scalar sector which our simplified model is a truncation of. 

It is our hope that both CMS and ATLAS continue their search for heavy Majorana neutrinos with masses above $1$ TeV early in run II. We have shown that in both simplified scenarios, where signal rates are small, there exist a series of interesting variables that can be used to probe lepton violating processes at the LHC. In addition, we encourage the collaborations to continue their efforts at determining the fake rate of jets into muons and electrons, especially at $13$ TeV where the effects of pile-up will be enormous.

\section*{Acknowledgements}

This work is supported in parts by the National Science and Engineering Council of Canada. W.P Pan is supported by Taiwan MOST, Grant No. 102-2112-M-007-014-MY3. The authors would like to thank Ayan Paul for his critical feedback throughout the progress of this work. J.N.N. and W.P Pan would  like to thank Prof. W.F. Chang for
discussions. J.N.N. also appreciates the kind hospitality of Prof. He of the National Center of Theoretical Science where part of this work was done.

%
%
%
\newpage
\appendix{\large{\bf{Appendices}}}

\section{T-channel amplitudes}\label{app:AppA}

In this section we analyze the properties of the $T$-channel amplitude in the Majorana neutrino extension of the SM. We separate the different contributions to the cross sections by looking at the different contributions from partons in the initial state and jets in the final state. The classification in terms of flavor and color and is gauge invariant. This is shown in Table~\ref{tab:TabAppA}.
\begin{table}[ht]\centering
 \tabcolsep 2.2 pt
\small
\begin{tabular}{|c||c|c|c|}
\hline
Type & Parton Pair & Jet Pair &  Color \\
\hline
\hline
$qq-qq-s$ & quark-quark: same flavor & quark-quark: same flavor & same color \\
\hline
$qq-qq-d$ & quark-quark: same flavor & quark-quark: same flavor & different color \\
\hline
$qQ-qq-s$ & quark-quark: different flavor & quark-quark: same flavor & same color \\
\hline
$qQ-qq-d$ & quark-quark: different flavor & quark-quark: same flavor & different color \\
\hline
$qq-qQ-s$ & quark-quark: same flavor & quark-quark: different flavor & same color \\
\hline
$qq-qQ-s$ & quark-quark: same flavor & quark-quark: different flavor & different color \\
\hline
$qQ-qQ$ & quark-quark: different flavor & quark-quark: different flavor &  \\
\hline
\hline
$qa-qa$ & quark-antiquark & quark-antiquark &  \\
\hline
$qa-qa-0$ & quark-antiquark & quark-antiquark & colorless \\
\hline
\end{tabular}
\caption{\small Classification of contributions to the $T$-channel mediated same-sign lepton production.} \label{tab:TabAppA}
\end{table}
The first two dominant contributions are $\sigma(uu\to dd~e^{+}e^{+}:qq-qq-d)$ and $\sigma(uu\to dd~e^{+}e^{+}:qq-qq-s)$, which are around the same order. The third and fourth dominant contributions are $\sigma(dd\to uu~e^{+}e^{+}:qq-qq-d)$ and $\sigma(dd\to uu~e^{+}e^{+}:qq-qq-s)$ respectively. The $qa-qa-0$ process can mix with the off-shell $S$-channel amplitude, but this turns out to be subdominant. Due to the hierarchy of the CKM matrix and the richness of the valance quarks in the proton, the four dominant processes are sufficient to perform a quantitative study of the $T$-channel amplitude. These are shown in Figure~\ref{fig:WWfusion}. Therefore we can write
\begin{eqnarray}
\sigma(uu\to dd~e^{+}e^{+}:qq-qq-d):N_{c}(N_{c}-1)\frac{1}{2}|M_{A}-M_{B}|^{2} \nonumber \\
\sigma(uu\to dd~e^{+}e^{+}:qq-qq-s):N_{c}\cdot\frac{1}{2}\cdot\frac{1}{2}|M_{A}-M_{B}-M_{C}+M_{D}|^{2}, \nonumber \\
\end{eqnarray}
where $N_{c}$ denotes the number of colors. To simplify the analysis we make the following definitions:
\begin{eqnarray}
\mathbb{C}_{A}&=&\frac{-1}{[M^{2}_{W}+2(p_{1}k_{1})][M^{2}_{W}+2(p_{2}k_{2})][M^{2}_{N_{R}}+2(p_{1}k_{1})+2(p_{1}l_{1})-2(l_{1}k_{1})]},\nonumber \\
\mathbb{C}_{B}&=&\frac{-1}{[M^{2}_{W}+2(p_{1}k_{1})][M^{2}_{W}+2(p_{2}k_{2})][M^{2}_{N_{R}}+2(p_{1}k_{1})+2(p_{1}l_{2})-2(l_{2}k_{1})]},\nonumber \\
\mathbb{C}_{C}&=&\frac{-1}{[M^{2}_{W}+2(p_{1}k_{2})][M^{2}_{W}+2(p_{2}k_{1})][M^{2}_{N_{R}}+2(p_{1}k_{2})+2(p_{1}l_{1})-2(l_{1}k_{2})]},\nonumber \\
\mathbb{C}_{D}&=&\frac{-1}{[M^{2}_{W}+2(p_{1}k_{2})][M^{2}_{W}+2(p_{2}k_{1})][M^{2}_{N_{R}}+2(p_{1}k_{2})+2(p_{1}l_{2})-2(l_{2}k_{2})]}.
\end{eqnarray}
Then the amplitude square for $qq-qq-d$ and $qq-qq-s$ can be written as
\begin{eqnarray}
|M|^{2}_{qq-qq-d}&=&8M^{2}_{N_{R}}g^{8}(k_{1}k_{2})\left((p_{1}l_{1})(p_{2}l_{2})[\mathbb{C}^{2}_{A}-\mathbb{C}_{A}\mathbb{C}_{B}]-(p_{1}l_{2})(l_{1}p_{2})[\mathbb{C}^{2}_{B}-\mathbb{C}_{A}\mathbb{C}_{B}]\right. \nonumber \\
&+&\left.(p_{1}p_{2})(l_{1}l_{2})\mathbb{C}_{A}\mathbb{C}_{B}\right), \nonumber \\
|M|^{2}_{qq-qq-s}&=&8M^{2}_{N_{R}}g^{8}(k_{1}k_{2})\left((p_{1}l_{1})(p_{2}l_{2}[(\mathbb{C}_{A}+\mathbb{C}_{C})^{2}-(\mathbb{C}_{A}+\mathbb{C}_{C})(\mathbb{C}_{B}+\mathbb{C}_{D})]\right. \nonumber \\
&+&\left.(p_{1}l_{2})(p_{2}l_{1}[(\mathbb{C}_{B}+\mathbb{C}_{D})^{2}-(\mathbb{C}_{A}+\mathbb{C}_{C})(\mathbb{C}_{B}+\mathbb{C}_{D})] \right. \nonumber \\
&+&\left.(p_{1}p_{2})(l_{1}l_{2}[(\mathbb{C}_{A}+\mathbb{C}_{C})(\mathbb{C}_{B}+\mathbb{C}_{D})] \right),
\end{eqnarray}
where we have used the massless approximation for jets, $(k_{1}+k_{2})^{2}=2(k_{1}k_{2})$. Both of the amplitudes are proportional to the invariant mass of the jet pair, and a stronger dependence comes from the Majorana mass. For large $M_{N_{R}}$, larger than the momentum transfer, $|q|^{2}$, the equations simplify since $\mathbb{C}_{A}\approx\mathbb{C}_{B}$ and $\mathbb{C}_{C}\approx\mathbb{C}_{D}$. In this limit the amplitudes reduce to
\begin{eqnarray}
 |M|^{2}_{qq-qq-d}&=&8M^{2}_{N_{R}}g^{8}(k_{1}k_{2})(p_{1}p_{2})(l_{1}l_{2})\mathbb{C}^{2}_{A},\nonumber \\
 |M|^{2}_{qq-qq-s}&=&8M^{2}_{N_{R}}g^{8}(k_{1}k_{2})(p_{1}p_{2})(l_{1}l_{2})(\mathbb{C}_{A}+\mathbb{C}_{C})^{2}.
\end{eqnarray}
Thus, in the limit where $\mathbb{C}_{A}=\mathbb{C}_{C}$, the two processes $qq-qq-d$ and $qq-qq-s$ yield the same contribution to the total cross section. In this region of parameter space, for $M^{2}_{N_{R}}>|q|^{2}$, the amplitudes are proportional to the same-sign lepton invariant mass. We make use of this variable to enhance a $T$-channel dominated signal form the SM background.
\section{The $\mathbb{E}$ $4$-particle kinematical variable}\label{app:AppB}
The $4$-particle final state variable introduced in Section~\ref{subsec:LRchannel} is defined by
\begin{equation}
\mathbb{E}=\epsilon_{\mu\nu\sigma\rho}P^{\mu}_{1}P^{\nu}_{2}P^{\sigma}_{3}P^{\rho}_{4},
\end{equation}
where $P_{1,2,3,4}$ denote the four-momentum of each of the particles in the final state. This variable can be significantly simplified in the rest frame defined by the $P_{2}+P_{3}+P_{4}$ invariant mass and after assigning the $z$-axis to the direction of $\vec{p}_{1}$. In this coordinate system, the $4$-momentum can be written as
\begin{equation}
P_{1}=|\vec{p}_{1}|\left(\begin{array}{lr}
1\\
0\\
0\\
1
\end{array}\right)~~~P_{2}=|\vec{p}_{2}|\left(\begin{array}{lr}
1\\
\sin\alpha\cos\beta\\
\sin\alpha\sin\beta\\
\cos\alpha
\end{array}\right),~~~P_{3}=|\vec{p}_{3}|\left(\begin{array}{lr}
1\\
\sin\sigma\cos\rho\\
\sin\sigma\sin\rho\\
\cos\sigma
\end{array}\right),~~~P_{2}+P_{3}+P_{4}=\left(\begin{array}{lr}
m_{2,3,4}\\
0\\
0\\
0\\
\end{array}\right),\\
\end{equation}
where $m_{2,3,4}$ denotes the invariant mass of particles $2,3$ and $4$, and the angles defined above parametrized the direction of the three-momentum with respect to the coordinate system. Therefore, with this choice, $\mathbb{E}$ can be written as
\begin{equation}
\mathbb{E}=-m_{2,3,4}~|\vec{p}_{1}||\vec{p}_{2}||\vec{p}_{3}|\sin\alpha\sin(\beta-\rho)\sin\sigma.
\end{equation}
and represents the invariant mass of the particles $1-3$ multiplied by the volume spanned by in three-momentum space. This variable has a large impact when the SU(2)$_{\text{R}}$ gauge boson is on-shell. To see this, we define $P_{1}=P_{l_{1}}$, $P_{2}=P_{l_{2}}$,  $P_{3}=P_{j_{1}}$ and $P_{4}=P_{j_{2}}$ in the massless limit to obtain $m_{1,2,3,4}=M_{W_{R}}$. If we choose the limit where $M_{N_{R}}<M_{W_{R}}$, then we further have $m_{l_{2}j_{1}j_{2}}=M_{N_{R}}$. With this in mind we can write
\begin{eqnarray}
\mathbb{E}_{M_{N_{R}}<M_{W_{R}}}&=&-\frac{M^{2}_{jj}(M^{2}_{W_{R}}-M^{2}_{N_{R}})(M^{2}_{N_{R}}-M^{2}_{jj})\sin\alpha\sin(\beta-\rho)\sin\sigma}{4[(M^{2}_{N_{R}}+M^{2}_{jj})+(M^{2}_{N_{R}}-M^{2}_{jj})(\cos\alpha\cos\sigma+\sin\alpha\sin\sigma\cos(\beta-\rho))}, \nonumber \\
\mathbb{E}_{M_{N_{R}}>M_{W_{R}}}~(a)&=&-\frac{M^{2}_{jj}(M^{2}_{W_{R}}-M^{2}_{l_{2}j_{1}j_{2}})(M^{2}_{l_{2}j_{1}j_{2}}-M^{2}_{jj})\sin\alpha\sin(\beta-\rho)\sin\sigma}{4[(M^{2}_{l_{2}j_{1}j_{2}}+M^{2}_{jj})+(M^{2}_{l_{2}j_{1}j_{2}}-M^{2}_{jj})(\cos\alpha\cos\sigma+\sin\alpha\sin\sigma\cos(\beta-\rho))}, \nonumber \\
\mathbb{E}_{M_{N_{R}}>M_{W_{R}}}~(b)&=&-\frac{M^{2}_{W_{R}}(M^{2}_{l_{1}l_{2}j_{1}j_{2}}-M^{2}_{N_{R}})(M^{2}_{N_{R}}-M^{2}_{W_{R}})\sin\alpha\sin(\beta-\rho)\sin\sigma}{4[(M^{2}_{N_{R}}+M^{2}_{W_{R}})+(M^{2}_{N_{R}}-M^{2}_{W_{R}})(\cos\alpha\cos\sigma+\sin\alpha\sin\sigma\cos(\beta-\rho))}, \nonumber \\
\end{eqnarray}
with scenarios $(a)$ and $(b)$ defined in Section~\ref{subsec:LRchannel}. This kinematical variable provides us with a nice separation between the SM background and heavy beyond-the-SM particles with masses above a TeV.



\end{document}